\documentclass[twocolumn,showpacs,aps,prd]{revtex4}

\usepackage{graphicx}
\usepackage{dcolumn}
\usepackage{amsmath}
\usepackage{epsfig}
\usepackage{color}

\input babarsym

\newcommand{\Ks}          {\KS}

\newcommand{\sysstat}{\ensuremath{\mathrm{(stat+syst)}}\xspace}

\newcommand{\CPp}	{\ensuremath{\CP\!+}\xspace}
\newcommand{\CPm}	{\ensuremath{\CP\!-}\xspace}
\newcommand{\CPpm}	{\ensuremath{\CP\!\pm}\xspace}

\newcommand{\Rkppm}	{\ensuremath{R_{K/\pi}^{\pm}}\xspace}

\newcommand{\Rkpp}	{\ensuremath{R_{K/\pi}^+}\xspace}
\newcommand{\Rkpm}	{\ensuremath{R_{K/\pi}^-}\xspace}
\newcommand{\Rkp}	{\ensuremath{R_{K/\pi}}\xspace}
\newcommand{\Rcp}	{\ensuremath{R_{\CP}}\xspace}
\newcommand{\Acp}	{\ensuremath{A_{\CP}}\xspace}
\newcommand{\Rcpp}	{\ensuremath{R_{\CP\!+}}\xspace}
\newcommand{\Rcpm}	{\ensuremath{R_{\CP\!-}}\xspace}
\newcommand{\Rcppm}	{\ensuremath{R_{\CP\!\pm}}\xspace}
\newcommand{\Acpp}	{\ensuremath{A_{\CP\!+}}\xspace}
\newcommand{\Acpm}	{\ensuremath{A_{\CP\!-}}\xspace}
\newcommand{\Acppm}	{\ensuremath{A_{\CP\!\pm}}\xspace}

\def\bb{\ensuremath{b\overline b}\xspace}

\newcommand{\dz}          {\Dz}

\newcommand{\dztokspiz}   {\ensuremath{\Dz\to\Ks\piz}\xspace}
\newcommand{\dztoksphi}   {\ensuremath{\Dz\to\Ks\phi}\xspace}
\newcommand{\dztoksomega} {\ensuremath{\Dz\to\Ks\omega}\xspace}
\newcommand{\dztokpi}     {\ensuremath{\Dz\to\Km\pip}\xspace}

\newcommand{\kk}	{\ensuremath{\Kp \kern -0.16em \Km}\xspace}
\newcommand{\kspiz}	{\ensuremath{\KS \piz}\xspace}
\newcommand{\ksphi}	{\ensuremath{\KS \phi}\xspace}
\newcommand{\ksomega}	{\ensuremath{\KS \omega}\xspace}
\newcommand{\kpi}	{\ensuremath{K^-\pi^+}\xspace}

\newcommand{\pipipiz}	{\ensuremath{\pi^-\pi^+\pi^0}\xspace}

\newcommand{\kknc}	{\ensuremath{KK}\xspace}		
\newcommand{\pipinc}	{\ensuremath{\pi\pi}\xspace}
\newcommand{\kspiznc}	{\ensuremath{\KS\piz}\xspace}
\newcommand{\ksphinc}	{\ensuremath{\KS\phi}\xspace}
\newcommand{\ksomeganc}	{\ensuremath{\KS\omega}\xspace}
\newcommand{\kpinc}	{\ensuremath{K\pi}\xspace}

\newcommand{\qqb}         {\ensuremath{q\overline q}\xspace}
\newcommand{\qq}          {\qqb}

\newcommand{\sigk}	  {\ensuremath{{\rm sig}(K)}\xspace}
\newcommand{\sigp}	  {\ensuremath{{\rm sig}(\pi)}\xspace}

\newcommand{\btodk}       {\ensuremath{B \to DK}\xspace}
\newcommand{\btdk}        {\btodk}

\newcommand{\btodp}       {\ensuremath{B \to D\pi}\xspace}
\newcommand{\btdp}        {\btodp}

\newcommand{\bmtodzk}     {\ensuremath{\Bm \to \Dz \Km}\xspace}

\newcommand{\bmtodcpk}    {\ensuremath{\Bm \to D_{\CP\pm} \Km}\xspace}
\newcommand{\bmtodcpmk}   {\ensuremath{\Bm \to D_{\CP-} \Km}\xspace}
\newcommand{\bmtodcppk}   {\ensuremath{\Bm \to D_{\CP+} \Km}\xspace}

\newcommand{\bmtodzp}     {\ensuremath{\Bm \to \Dz \pim}\xspace}

\newcommand{\bmtodcpp}	  {\ensuremath{\Bm \to D_{\CP\pm} \pim}\xspace}

\newcommand{\bptodzbk}	  {\ensuremath{\Bp \to \Dzb \Kp}\xspace}
\newcommand{\bptodcpk}	  {\ensuremath{\Bp \to D_{\CP\pm} \Kp}\xspace}
\newcommand{\bptodcpmk}   {\ensuremath{\Bp \to D_{\CP-} \Kp}\xspace}
\newcommand{\bptodcppk}   {\ensuremath{\Bp \to D_{\CP+} \Kp}\xspace}

\newcommand{\bptodzbp}	  {\ensuremath{\Bp \to \Dzb \pip}\xspace}
\newcommand{\bptodcpp}	  {\ensuremath{\Bp \to D_{\CP\pm}\pip}\xspace}

\newcommand{\de}	{\DeltaE}

\newcommand{\deshift}	{\ensuremath{\DeltaE_{\tt shift}}\xspace}
\newcommand{\dep}	{\ensuremath{\DeltaE(\pi)}\xspace}
\newcommand{\dek}	{\ensuremath{\DeltaE(K)}\xspace}
\newcommand{\fshr}	{\ensuremath{\mathcal{F}}\xspace}

\newcommand{\bfi}	{\begin{figure}}
\newcommand{\efi}	{\end{figure}}
\newcommand{\beq}	{\begin{equation}}
\newcommand{\eeq}	{\end{equation}}
\newcommand{\beqa}	{\begin{eqnarray}}
\newcommand{\eeqa}	{\end{eqnarray}}

\def\sss{\scriptscriptstyle}
\def\barpd{{\raise.35ex\hbox{${\sss (}$}}--{\raise.35ex\hbox{${\sss )}$}}}
\def\dbarp{\hbox{$D^{0}$\kern-1.3em\raise1.5ex\hbox{\barpd}}\:}
\def\dstarbarp{\hbox{$D^{*0}$\kern-1.5em\raise1.5ex\hbox{\barpd}}\:}

\long\def\blfootnote[#1]#2{\begingroup%
\def\thefootnote{\fnsymbol{footnote}}\footnote[#1]{#2}\endgroup}

\newcommand{\BABARPubYear}    {10}
\newcommand{\BABARPubNumber}  {008}
\newcommand{\SLACPubNumber} {14187}
\newcommand{\LANLNumber}    {1007.0504}

\def\AcppVal{0.25}
\def\AcppErrStat{0.06}
\def\AcppErrSyst{0.02}
\def\AcpmVal{-0.09}
\def\AcpmErrStat{0.07}
\def\AcpmErrSyst{0.02}
\def\RcppVal{1.18}
\def\RcppErrStat{0.09}
\def\RcppErrSyst{0.05}
\def\RcpmVal{1.07}
\def\RcpmErrStat{0.08}
\def\RcpmErrSyst{0.04}

\def\CORstat{\begin{eqnarray}
\label{eq:cor_stat}
C_{\stat}[\vec{y}]
= \left( \begin{array}{cccc}
1    & 0.0  & -0.08  &  0.0  \\
     & 1    &  0.0   &  0.03 \\
     &      &  1     &  0.10 \\
     &      &        &  1              
\end{array} \right).
\end{eqnarray}}

\def\CORsyst{\begin{equation}
\label{eq:cor_syst}
C_{\syst}[\vec{y}]
= \left( \begin{array}{cccc}
1    & 0.56 & -0.06  &  0     \\
     & 1    &  0     &  0     \\
     &      &  1     &  0.13  \\
     &      &        &  1	       
\end{array} \right).
\end{equation}}

\def\AcpmValNoKsPhi{-0.08}
\def\AcpmErrStatNoKsPhi{0.07}
\def\AcpmErrSystNoKsPhi{0.02}
\def\RcpmValNoKsPhi{1.03}
\def\RcpmErrStatNoKsPhi{0.09}
\def\RcpmErrSystNoKsPhi{0.04}

\def\CORstatNoKsphi{\begin{eqnarray}
\label{eq:cor_stat_noksphi}
C_{\stat}[\vec{y}]
= \left( \begin{array}{cccc}
1    & 0    & -0.08  &  0    \\
     & 1    &  0     &  0.04 \\
     &      &  1     &  0.09 \\
     &      &        &  1              
\end{array} \right)\ ,
\end{eqnarray}}

\def\CORsystNoKsphi{\begin{eqnarray}
\label{eq:cor_syst_noksphi}
C_{\syst}[\vec{y}]
= \left( \begin{array}{cccc}
1    & 0.56  & -0.06  &  0    \\
     & 1     &  0     &  0    \\
     &       &  1     &  0.12 \\
     &       &        &  1              
\end{array} \right).
\end{eqnarray}}

\def\dgLoAi{11.3}
\def\dgHiAi{22.7}
\def\dgLoAj{80.8}
\def\dgHiAj{99.2}
\def\dgLoAk{157.3}
\def\dgHiAk{168.7}

\def\dgLoBi{7.0}
\def\dgHiBi{173.0}
\def\dgLoBj{187.0}
\def\dgHiBj{353.0}

\def\rLoA{0.24}
\def\rHiA{0.45}

\def\rLoB{0.06}
\def\rHiB{0.51}

\def\rbVal{0.35}
\def\rbErrHi{+0.10}
\def\rbErrLo{-0.11}

\def\xpValNoKsPhi{-0.057}
\def\xpErrStatNoKsPhi{0.039}
\def\xpErrSystNoKsPhi{0.015}

\def\xmValNoKsPhi{0.132}
\def\xmErrStatNoKsPhi{0.042}
\def\xmErrSystNoKsPhi{0.018}

\def\rbSqValNoKsPhi{0.105}
\def\rbSqErrStatNoKsPhi{0.067}
\def\rbSqErrSystNoKsPhi{0.035}

\def\ARARval{0.19}
\def\ARARerr{0.11}

\def\sigStat{3.7}
\def\sigStatSyst{3.6}

\def\figurebox#1#2#3{
    \def\arg{#3}
    \ifx\arg\empty
    {\hfill\vbox{\hsize#2\hrule\hbox to #2{\vrule\hfill\vbox to #1{\hsize#2\vfill}\vrule}\hrule}\hfill}
    \else
    {\hfill\epsfbox{#3}\hfill}
    \fi}

\begin{document}

\begin{minipage}[t]{\textwidth}
  \begin{minipage}[t]{0.49\textwidth}
    \begin{flushleft}
    \babar-PUB-\BABARPubYear/\BABARPubNumber\\
    SLAC-PUB-\SLACPubNumber\\
    arXiv:\LANLNumber\ [hep-ex]\\
    \end{flushleft}
  \end{minipage}
\end{minipage}

\vspace{1.5cm}

\preprint{\babar-PUB-\BABARPubYear/\BABARPubNumber} 
\preprint{SLAC-PUB-\SLACPubNumber} 

\title{\boldmath \large \bf Measurement of \CP observables in $B^\pm\to D_{\CP}K^\pm$ decays and constraints on the CKM angle $\gamma$}

%
\author{P.~del~Amo~Sanchez}
\author{J.~P.~Lees}
\author{V.~Poireau}
\author{E.~Prencipe}
\author{V.~Tisserand}
\affiliation{Laboratoire d'Annecy-le-Vieux de Physique des Particules (LAPP), Universit\'e de Savoie, CNRS/IN2P3,  F-74941 Annecy-Le-Vieux, France}
\author{J.~Garra~Tico}
\author{E.~Grauges}
\affiliation{Universitat de Barcelona, Facultat de Fisica, Departament ECM, E-08028 Barcelona, Spain }
\author{M.~Martinelli$^{ab}$}
\author{A.~Palano$^{ab}$ }
\author{M.~Pappagallo$^{ab}$ }
\affiliation{INFN Sezione di Bari$^{a}$; Dipartimento di Fisica, Universit\`a di Bari$^{b}$, I-70126 Bari, Italy }
% ...
\collaboration{The \babar\ Collaboration}
\noaffiliation

\date{\today}

\begin{abstract}
Using the entire sample of 467 million $\FourS \to \BB$
decays collected with the \babar\ detector at the \pep2
asymmetric-energy $B$ factory at SLAC, we perform an analysis
of $B^\pm \to D K^\pm$ decays, using decay modes in which the 
neutral $D$ meson decays to either \CP-eigenstates or non-\CP-eigenstates.
We measure the partial decay rate charge asymmetries for
\CP-even and \CP-odd $D$ final states to be
$\Acpp = \AcppVal\pm\AcppErrStat\pm\AcppErrSyst$ and 
$\Acpm = \AcpmVal\pm\AcpmErrStat\pm\AcpmErrSyst$, respectively, where
the first error is the statistical and the second is
the systematic uncertainty.
The parameter \Acpp is different from zero with a significance of \sigStatSyst\
standard deviations, constituting evidence for direct \CP
violation. We also measure the ratios of the charged-averaged $B$
partial decay rates in \CP and non-\CP decays,
$\Rcpp = \RcppVal\pm\RcppErrStat\pm\RcppErrSyst$ and 
$\Rcpm = \RcpmVal\pm\RcpmErrStat\pm\RcpmErrSyst$.
We infer frequentist confidence intervals for the angle $\gamma$ of the
unitarity triangle, for the strong phase difference $\delta_B$, and for the
amplitude ratio $r_B$, which are related to the $B^-\to DK^-$ decay
amplitude by $r_Be^{i(\delta_B-\gamma)} = A(B^-\to \Dzb K^-)/A(B^-\to\Dz
K^-)$.
Including statistical and systematic uncertainties, we obtain 
$\rLoA<r_B<\rHiA$ ($\rLoB<r_B<\rHiB$) and, modulo $180\degrees$,
$\dgLoAi\degrees<\gamma<\dgHiAi\degrees$ or 
$\dgLoAj\degrees<\gamma<\dgHiAj\degrees$ or
$\dgLoAk\degrees<\gamma<\dgHiAk\degrees$
($\dgLoBi\degrees<\gamma<\dgHiBi\degrees$) 
at the 68\% (95\%) confidence level.
\end{abstract}

\pacs{13.25.Hw, 12.15.Hh, 11.30.Er}

\maketitle

\section{Introduction}
\label{sec:introduction}

In the standard model (SM) of fundamental particles,
\CP violation in weak interactions is allowed
by a single, irreducible phase in the $3\times 3$ 
Cabibbo-Kobayashi-Maskawa (CKM) quark flavor-mixing
matrix~\cite{cabibbo,ckmmatrix}.
The unitarity of the CKM matrix, $V$,
implies a set of relations among its elements $V_{ij}$,
in particular the condition $V_{ud}^{}V_{ub}^* +V_{cd}^{}V_{cb}^*
+V_{td}^{}V_{tb}^* = 0$, which can be depicted in the complex
plane as a ``unitarity'' triangle, whose sides and angles 
are related to the magnitudes and phases of the six elements $V_{id}$
and $V_{ib}$, where $i=u,c,t$.
Overconstraining the unitarity triangle by means of precise
measurements of all its sides and angles allows tests of whether the CKM
mechanism is the correct description of \CP violation. Any
inconsistencies among the various experimental constraints would reveal
effects of physics beyond the standard model.

After a decade of successful operation and a total of about 1.3 billion
\BB pairs collected by the \babar\ and Belle experiments, the three CKM
angles have been measured with varied precision. The angle $\beta$ has
been measured with the highest precision, to around $1\degrees$, using 
$B^0\to (c\bar{c})K^{(*)0}$ decays.
Using a variety of two-body $B$ decays ($B\to\pi\pi,\ \rho\pi,\ \rho\rho$ and
$a_1(1260)\pi$)
the angle $\alpha$ has been measured to a precision
of around $4\degrees$. The
angle $\gamma$ has a relatively large uncertainty, around
$14\degrees$, compared with $\alpha$ and
$\beta$. The lack of precision in our knowledge of $\gamma$ reflects the
difficulty in measuring this angle. The uncertainties of the CKM
angles quoted in this paragraph are taken from~\cite{CKMfitter}.

Several techniques for measuring $\gamma$ in a theoretically clean way are
based on $B$ meson decays to open-charm final states,
$D^{(*)0}X_s$ and $\Dbar^{(*)0}X_s$ ($X_s = K^{(*)\pm},\
K^{(*)0}$). In these decays, the interference between the $b\to c\bar u
s$ and $b\to u\bar c s$ tree amplitudes, when the \Dz and
\Dzb decay to a common final state, leads to observables
that depend on the relative weak phase $\gamma$.
The size of the interference also depends on the magnitude of the ratio
$r_B$ and the relative phase strong phase $\delta_B$ of the two
amplitudes, which can not be precisely calculated from theory.
They can be extracted directly from data by
simultaneously reconstructing several related $B\to DK$ decays. Many methods
have been proposed to extract $\gamma$ from $B$
decays using $D^{(*)}K^{(*)\pm}$ and $D^{(*)}K^{(*)0}$ final states (here and 
in the following $D$ refers to any admixture of the neutral \Dz 
meson and its \CP-conjugate \Dzb).
The three methods that have been used most productively to date are
the ``GLW'' method~\cite{GLW1,GLW2}, based on Cabibbo-suppressed
$D$ decays to \CP-eigenstates, such as $K^+K^-$ or $K^0_S\pi^0$; the ``ADS''
method~\cite{ADS1,ADS2}, where the $D$ is reconstructed in
Cabibbo-favored and doubly-Cabibbo-suppressed final states such 
as $K^\pm\pi^\mp$; and the ``GGSZ'' method~\cite{DALITZ}, which studies the 
Dalitz-plot distribution of the products of $D$ decays to multi-body
self-conjugate final states, such as $K^0_S\pi^+\pi^-$.
A common problem with these methods is the small overall branching
fraction of these decays ranging from $5\times 10^{-6}$ to $5\times
10^{-9}$. Therefore a precise determination of $\gamma$ requires a very
large data sample.
\babar\ has published
several $\gamma$ related measurements: 
GLW analyses of $B^\pm\to DK^\pm$~\cite{babar_d0k_GLW_PRD}, 
$D^*K^\pm$~\cite{babar_dstar0k_GLW_PRD} and $DK^{*\pm}$~\cite{babar_d0kstar_GLWADS_PRD} decays; ADS analyses of $B^\pm\to
D^{(*)}K^\pm$~\cite{babar_dk_kpi_ADS_PRD,babar_dk_kpipi0_ADS_PRD},
$DK^{*\pm}$~\cite{babar_d0kstar_GLWADS_PRD} and $B^0\to DK^{*0}$
\cite{babar_d0kstar0_ADS_PRD}; and GGSZ analyses of
$B^\pm\to D^{(*)}K^{(*)\pm}$~\cite{babar_dk_DALITZ_PRD,babar_dk_DALITZ_PRL} and $B^0\to DK^{*0}$
decays~\cite{babar_d0kstar0_DALITZ_PRD}. To date, the single most precise
experimental determination of $\gamma$ from \babar\ is $\gamma = 
(68\pm 14\pm 4\pm 3)\degrees$ and
$39\degrees<\gamma<98\degrees$,
obtained from the GGSZ analysis of $B^\pm\to
D^{(*)}K^{(*)\pm}$ decays~\cite{babar_dk_DALITZ_PRL}. In this measurement, the first error represents the 
statistical uncertainty, the second is the experimental systematic 
uncertainty, and the third reflects the uncertainty on the description of 
the $D$ Dalitz-plot distributions. 

\section{\boldmath GLW analysis of \btodk decays}
\label{sec:glw_introduction}

In this paper we present the update of the GLW
analysis of $B^\pm\to DK^\pm$ decays based on the full
\babar\ dataset collected near the \FourS resonance. 
In addition to a 22\% increase in statistics of the data
sample, this study benefits from other
significant improvements compared to our previous
result~\cite{babar_d0k_GLW_PRD}:
\begin{itemize}
\item More refined charged track reconstruction and particle identification
  algorithms, with higher purity and efficiency, have been employed;
\item The event shape variable \fshr, used to discriminate the signal 
  from the continuum $e^+e^-\to q\bar{q}$ background (described in
  detail in Section~\ref{sec:selection}) has been removed from the
  selection criteria and has instead been included in the final fit to the
  selected $B$ candidates. This allows us to increase the signal
  efficiency by about 30\%. At the same time it provides a larger
  sample of continuum background events, thus allowing for the
  determination of the background properties directly from data (see
  Section~\ref{sec:fit});
\item Better kaon/pion separation, which is needed to distinguish 
  $B^\pm\to DK^\pm$
  candidates from the twelve times more abundant $B^\pm\to D\pi^\pm$ decays, is 
  achieved through the use of a global likelihood based not only on
  the Cherenkov angle $\theta_C$ reconstructed by the 
  Cherenkov detector, but also on the specific energy loss 
  ${\rm d}E/{\rm d}x$ measured by the tracking devices. The inclusion of
  ${\rm d}E/{\rm d}x$ in the likelihood increases the
  kaon identification efficiency and decreases the pion misidentification
  both at low momentum and outside of the geometrical 
  acceptance of the Cherenkov detector (which is 10\% lower than the 
  acceptance of the tracking devices).
\end{itemize}

In order to determine $\gamma$ from $B^\pm\to DK^\pm$ decays
with the GLW method, we measure the two direct-\CP-violating
partial decay rate asymmetries,
\begin{equation}
  \Acppm \equiv
  \frac{\Gamma(\bmtodcpk)-\Gamma(\bptodcpk)}{\Gamma(\bmtodcpk)+\Gamma(\bptodcpk)}\,,
  \label{eq:def_Acp}
\end{equation}
and the two ratios of charge averaged partial rates using $D$ decays
to \CP and flavor eigenstates,
\begin{equation}
  \Rcppm \equiv 2 \,
  \frac{\Gamma(\bmtodcpk)+\Gamma(\bptodcpk)}{\Gamma(\bmtodzk)+\Gamma(\bptodzbk)}\,,
  \label{eq:def_Rcp} 
\end{equation}
where $D_{\CPpm}$ refer to the \CP eigenstates of the $D$ meson system.
We then extract $\gamma$, together with the other two unknowns $r_B$
and $\delta_B$, by means of a frequentist procedure, which exploits
the following relations~\cite{GLW1,GLW2}, neglecting \Dz--\Dzb mixing~\cite{MIXING}:
\begin{eqnarray}
	\label{eq:rcp}
	R_{\CP\pm} &=& 1+r_B^2 \pm 2r_B\cos\delta_B \cos\gamma\,, \\
	\label{eq:acp}
	A_{\CP\pm} &=& \frac{\pm 2 r_B \sin\delta_B \sin\gamma}{1+r_B^2 \pm2r_B\cos\delta_B \cos\gamma}.
\end{eqnarray}
Here, $r_B{\equiv}\left|A(B^-{\to}{\Dzb}K^-)/A(B^-{\to}{\Dz}K^-)\right|$
is the magnitude of the ratio of the amplitudes for $B^-{\to}{\Dzb}K^-$
and $B^-{\to}{\Dz}K^-$ and $\delta_B$ the difference of their strong
phases. Taking into account the CKM factor ($|V_{ub}V_{cs}/V_{cb}V_{us}|
\approx 0.4$) and  color-suppression of the $B^-{\to}{\Dzb}K^-$
amplitude, $r_B$ is expected to be around 0.1. The current world averages
for the $B^\pm\to DK^\pm$ GLW observables from the measurements
in~\cite{babar_d0k_GLW_PRD,belle_dk_GLW_PRD,CDF_d0k_GLW} are
summarized in Table~\ref{tab:wa_GLW}.
\begin{table}[!h]
\begin{center}
\caption{World averages at 68\% confidence level~\cite{HFAG} for the GLW observables in $B\to DK$ decays.} 
\label{tab:wa_GLW}
\begin{tabular}{lcc}
\hline\hline\\[-2.5ex]
\CP of the $D$ & $R_{\CP}$	  & $A_{\CP}$      \\
\hline
$+1$           & $1.10\pm0.09$  & $\phantom{-}0.24\pm0.07$ \\
$-1$           & $1.06\pm0.10$  & $-0.10\pm0.08$ \\
\hline\hline
\end{tabular}
\end{center}
\end{table}
The world averages for the parameters $r_B$ and $\delta_B$ are
$r_B{=}0.104^{+0.015}_{-0.025}$ and
$\delta_B{=}(117^{+17}_{-24})\degrees$ at 68\% confidence level (CL)~\cite{CKMfitter}.

To reduce the systematic uncertainties from branching fractions
and reconstruction efficiencies of different $D$ channels appearing
in the numerator and denominator of Eq.~\ref{eq:def_Rcp}, we approximate
$R_{\CPpm}$ with the double ratios
\begin{equation}
  \label{eq:rpmdef}
  R_{\CP\pm} \approx \frac{\Rkppm}{\Rkp}\ ,
\end{equation}
where 
\begin{equation}
  \Rkppm \equiv
  \frac{\Gamma(\bmtodcpk)+\Gamma(\bptodcpk)}{\Gamma(\bmtodcpp)+\Gamma(\bptodcpp)}\,,
  \label{eq:def_Rkpi_cp}
\end{equation} 
and
\begin{equation}
  \Rkp \equiv
  \frac{\Gamma(\bmtodzk)+\Gamma(\bptodzbk)}{\Gamma(\bmtodzp)+\Gamma(\bptodzbp)}\,.
  \label{eq:def_Rkpi}
\end{equation} 
Equation~\ref{eq:rpmdef} would be exact in the limit in which 
the Cabibbo-suppressed contributions to the $B^\pm\to D\pi^\pm$ amplitudes 
vanish, as well as terms proportional to $r_B r_D\approx 5\times10^{-3}$, as we
will discuss in Section~\ref{sec:systematics}.
This approximation results in a systematic uncertainty on the final values of
$R_{\CP\pm}$. 

The paper is organized as follows. In Section~\ref{sec:detector}
we describe the data sample used for these measurements and
the main features of the \babar\ detector and of the PEP-II storage
rings. In Section~\ref{sec:selection} we
summarize the procedure adopted to select $B^\pm\to Dh^\pm$ candidates and
suppress the main backgrounds. In Section~\ref{sec:fit} we introduce the
simultaneous extended maximum likelihood fit used to extract
the observables \Rcppm and \Acppm. In Section~\ref{sec:peaking_bkg} we explain
how, by applying the same fit procedure to selected control samples,
we estimate the irreducible background present in the final
samples. A discussion of the sources of systematic
uncertainties and the evaluation of the uncertainties is presented in
Section~\ref{sec:systematics}. Section~\ref{sec:results} lists the
final results on the GLW observables $R_{\CP\pm}$ and $A_{\CP\pm}$,
including statistical and systematic uncertainties. It also contains a
description of the statistical method used to construct
frequentist confidence intervals for the 
parameters $\gamma$, $\delta_B$, and $r_B$. 
Section~\ref{sec:summary} gives a summary of our results.

\section{Data sample and Detector}
\label{sec:detector}

The measurements presented in this paper use the entire \BB data sample collected
with the \babar\ detector at the \pep2 asymmetric-energy $B$ factory at
the SLAC National Accelerator Laboratory. The \BB pairs are
produced from the decays of \FourS mesons that originate in
collisions of 9.0\gev electrons and 3.1\gev positrons ($\sqrt{s}{=}
10.58\gev{=}M_{\FourS}c^2)$. In total, $(467\pm 5)\times 10^6$ \BB
pairs, approximately equally divided into $\Bz\Bzb$ and $B^+B^-$,
have been collected in the years from 1999 until early 2008.
The $B$ meson pairs are produced almost at rest in the \FourS 
center-of-mass (CM) frame, but the asymmetric beam energies 
boost them in the laboratory frame by
$(\beta\gamma)_{\rm CM}\approx 0.56$.

The \babar\ detector is described in detail elsewhere~\cite{babar_detector}. 
Primary
and secondary vertex reconstruction and charged-particle tracking are
provided by a five layer double-sided  silicon vertex tracker and a
40 layer drift chamber. Charged particle identification (PID) is
provided by measurement of specific ionization energy loss in the
tracking devices and of the Cherenkov radiation cone in a
ring-imaging detector. Photons and electrons are identified by
combining the information from the tracking devices and 
the energy deposits in the electromagnetic calorimeter, which consists of 6580
thallium-doped CsI crystals. These systems are located inside a
$1.5\,\textrm{T}$ solenoidal superconducting magnet. Finally, the flux
return of the magnet is instrumented with resistive plate chambers and
limited streamer tubes in order to discriminate muons from pions. We use
the {\tt GEANT4}~\cite{geant} software toolkit to simulate interactions of
particles in the detector, taking into account the varying
accelerator and detector conditions.

\section{Event Selection}
\label{sec:selection}

We reconstruct $B^\pm \to Dh^\pm$ decays, where the charged
track $h$ is either a kaon or a pion. Neutral $D$ mesons are reconstructed 
in the \CP-even eigenstates $\pi^-\pi^+$ and $K^-K^+$ ($D_{\CPp}$), in
the \CP-odd eigenstates $\KS\pi^0$, $\KS\phi$ and $\KS\omega$
($D_{\CPm}$), and in the non-\CP-eigenstate $K^-\pi^+$ ($D^0$ from $B^-
\to D^0h^-$) or $K^+\pi^-$ ($\Dzb$ from $B^+ \to \Dzb h^+$).
\CP violation in the \Kz--\Kzb system is neglected, $i.e.$ the 
\KS is assumed to be a pure $\CP=+1$ eigenstate.
The $D_{\CP}$ daughters are reconstructed in the decay modes
$\KS\to\pi^+\pi^-$, $\phi\to K^+K^-$ and $\omega\to\pi^-\pi^+\pi^0$.

We optimize all our event selection requirements by maximizing the
significance of the expected $B^\pm\to D K^\pm$ signal yield, defined
as $N_{\rm sig}/\sqrt{N_{\rm sig}+N_{\rm bkg}}$, where $N_{\rm sig}$
($N_{\rm bkg}$) is the expected signal (background) yield.
The optimization is done for each $D$ decay channel using simulated
signal and background events, which are generated with the {\tt EVTGEN}
software package~\cite{evtgen}.

Neutral pions are reconstructed by combining pairs of photon candidates
with energy deposits larger than 30\mev that are not matched to charged
tracks and whose energy deposition profile is consistent with that
expected from a photon. The photon pair invariant mass is required to
differ from the nominal \piz mass~\cite{PDG2008} by less than 2.5 times
its resolution ($\sigma \approx 6 \mevcc$) and the total \piz energy in
the laboratory frame must be greater than 240\mev for $D\to\KS\piz$ and
210\mev for $\omega\to\pi^+\pi^-\piz$.

Neutral kaons are reconstructed from pairs of oppositely charged tracks
with invariant mass within $2.5\sigma$ ($\sigma \approx 2.1\mevcc$) of
the nominal \KS mass~\cite{PDG2008}. The ratio between the 
\KS signed 3-dimensional flight length and its uncertainty,
determined from the position of the \KS 
and the $D$ decay vertices and the \KS momentum direction, 
must be greater than 1.9, 2.0, and 2.2 for $D\to\KS\piz$,
$D\to\KS\phi$, and $D\to\KS\omega$, respectively.

The $\phi$ candidates are reconstructed from pairs of oppositely charged
tracks passing kaon identification criteria with typical
kaon selection efficiency of $\approx 98\%$ and pion misidentification of $\approx 15\%$. 
The
two tracks are assigned the kaon mass hypothesis and their invariant
mass is required to be within $6.5\mevcc$ of the nominal $\phi$ 
mass~\cite{PDG2008} (the resolution is $\sigma=1.0\mevcc$ and the natural width 
is $\Gamma_\phi=4.3\mev$).
We also require that the helicity angle $\theta_H$ between the flight
direction of one of the two kaons and the $D$ flight direction, in the
$\phi$ rest frame, satisfies the condition $|\cos\theta_H|>0.4$. This
requirement exploits the fact that in $D\to\KS\phi$ decays the $\phi$ is
produced in a longitudinally polarized state, thus $\cos\theta_H$
follows a $\cos^2\theta_H$ distribution, while in $\phi$ candidates not
from $D\to\KS\phi$ decays, $\cos\theta_H$ is approximately uniformly
distributed.

The $\omega$ candidates are reconstructed from $\pi^+\pi^-\pi^0$
combinations with invariant mass within $17\mevcc$
($2\Gamma_\omega$) of the nominal $\omega$ mass~\cite{PDG2008} 
(the resolution is $\sigma=6.9\mevcc$).
The charged pion candidates are required to pass pion
identification criteria with pion selection efficiency around 98\% and kaon
misidentification rate around 12\%. To improve the $\omega$ momentum
resolution, the invariant mass of the two photons forming the $\piz$
candidate is constrained to the nominal \piz mass. We define $\theta_N$ as
the angle between the normal to the $\omega$ decay plane and the $D$
momentum in the $\omega$ rest frame, and $\theta_{\pi\pi}$ as the angle
between the flight direction of one of the three pions in the $\omega$
rest frame and the flight direction of one of the other two pions in the
two-pion rest frame. The quantities $\cos\theta_N$ and
$\cos\theta_{\pi\pi}$ follow $\cos^2\theta_N$ and
$(1-\cos^2\theta_{\pi\pi})$ distributions, respectively, for the signal,
and are almost uniformly distributed for wrongly reconstructed $\omega$
candidates. We require the product $\cos^2\theta_N\sin^2\theta_{\pi\pi}
> 0.046$.

Neutral $D$ candidates are formed from two-body combinations
of $K^\pm$, $\pi^\pm$, \KS, \piz, $\phi$ and $\omega$ candidates
consistent with one of the six $D$ decay channels under study.  To
improve the $D_{\CPm}$ momentum resolution, the invariant masses of the
\piz and \KS daughters
are constrained to the nominal \piz and \KS masses. To suppress poorly
reconstructed $D$ candidates and candidates from random combinations, we
perform a geometric fit of the $D$ daughters to a common origin, and
reject $D$ candidates for which the $\chi^2$ probability of the vertex
fit is lower than 0.01\%. The invariant mass of a $D$ candidate $M_D$
must be within  a range that corresponds to slightly more than twice
the $M_D$ resolution, which varies from about 6\mevcc for the \ksphi
channel to about 44\mevcc for the \kspiz channel.
We apply the following particle identification criteria to the charged daughters of the $D$ meson:
in $D\to\pipi$, the two pion candidates must pass the same
pion identification criteria adopted in the reconstruction of
$\omega\to\pipipiz$; in $D\to\kk$, the two kaon candidates are required
to pass tighter kaon identification criteria than those applied to the
$\phi$ daughters (typical kaon selection efficiency around 94\%, and pion
misidentification rate around 6\%); in $D\to\kpi$, the kaon candidate
must pass the same kaon identification criteria required for the $\phi$
daughters. 
In order to reduce the large combinatorial background from random
combinations of tracks and photons in $e^+e^-\to q\bar{q}$ events 
($q=u,d,s,c$), we put requirements on the cosine of the $D$ decay angle,
$|\cos\theta_D|$. We define $\theta_D$ as the angle between one of the
$D$ daughters in the $D$ rest frame, and the direction of the $D$ meson
in the $B$ rest frame. Due to angular momentum conservation we expect
the distribution of $\cos\theta_D$ to be uniform for $B^\pm\to Dh^\pm$,
$D\to\pipi$ and $D\to\KS\piz$ signal events, while for \qq events the
distribution is strongly peaked  at $\pm 1$. We require
$|\cos\theta_D|<0.74$ (0.99) for the $B^\pm\to Dh^\pm$, $D\to\pipi$
($D\to\kspiz$) channel.

\begin{figure*}[!htb]
\begin{center}
\includegraphics[width=0.39\textwidth]{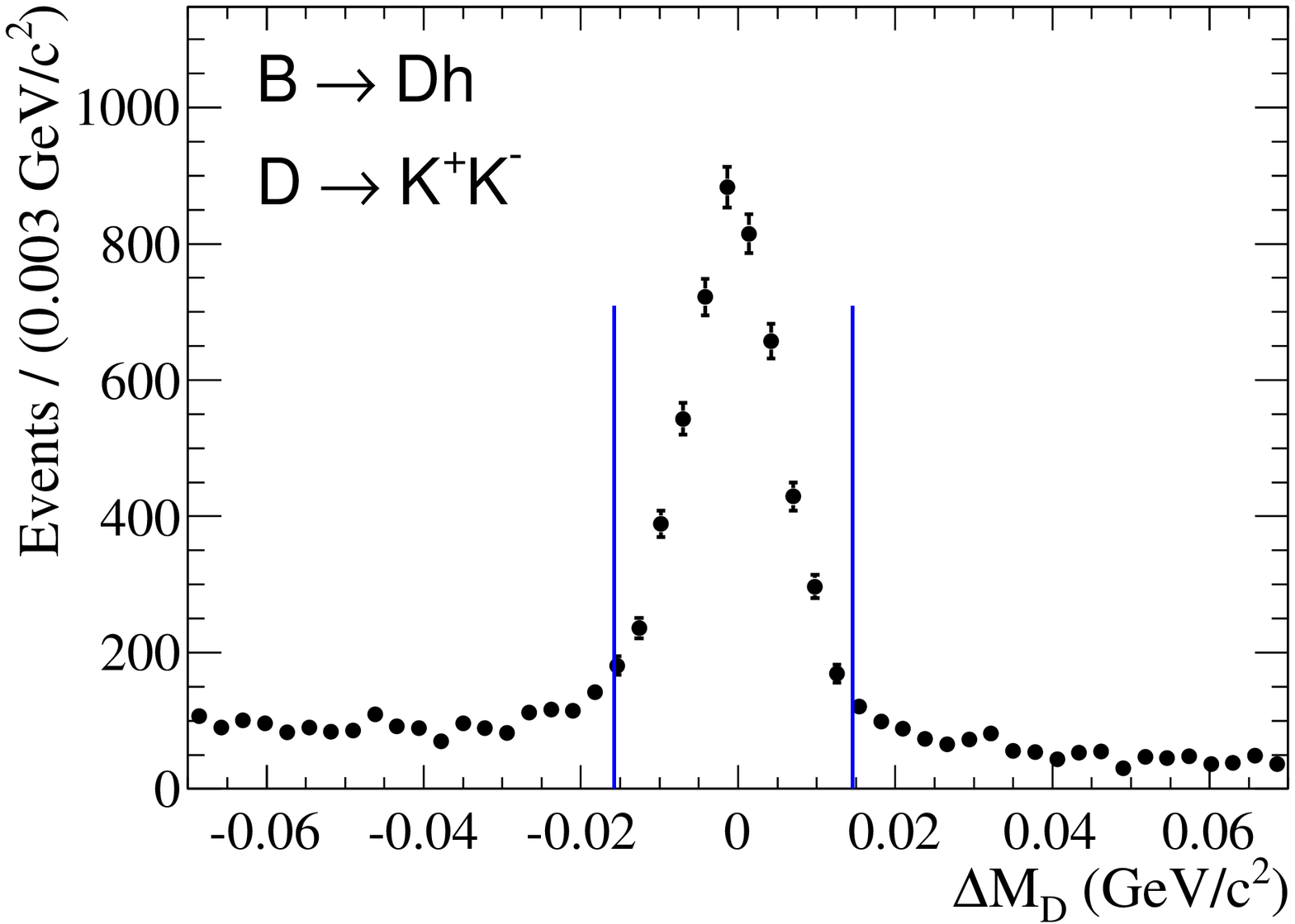}
\includegraphics[width=0.39\textwidth]{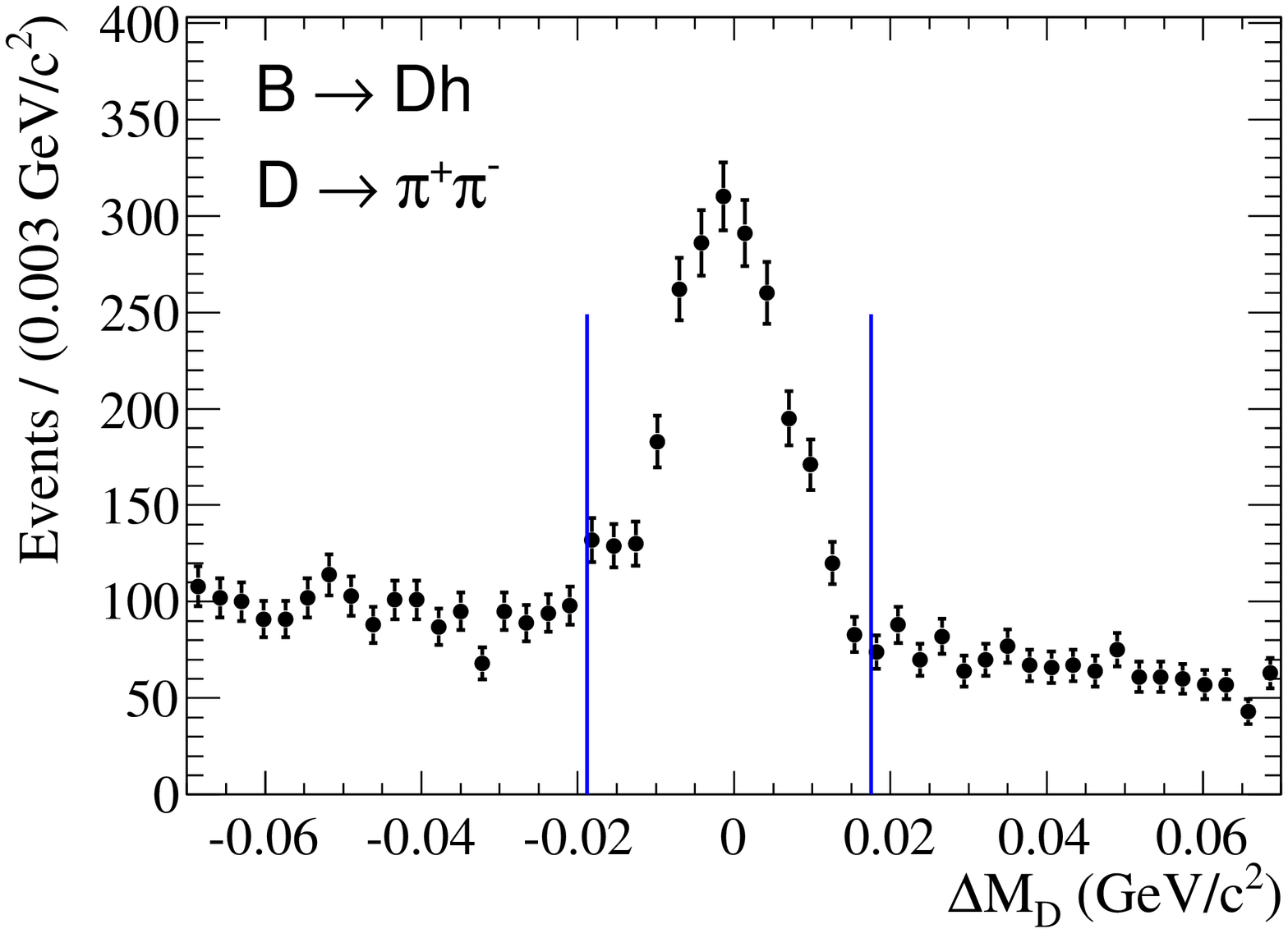}
\includegraphics[width=0.39\textwidth]{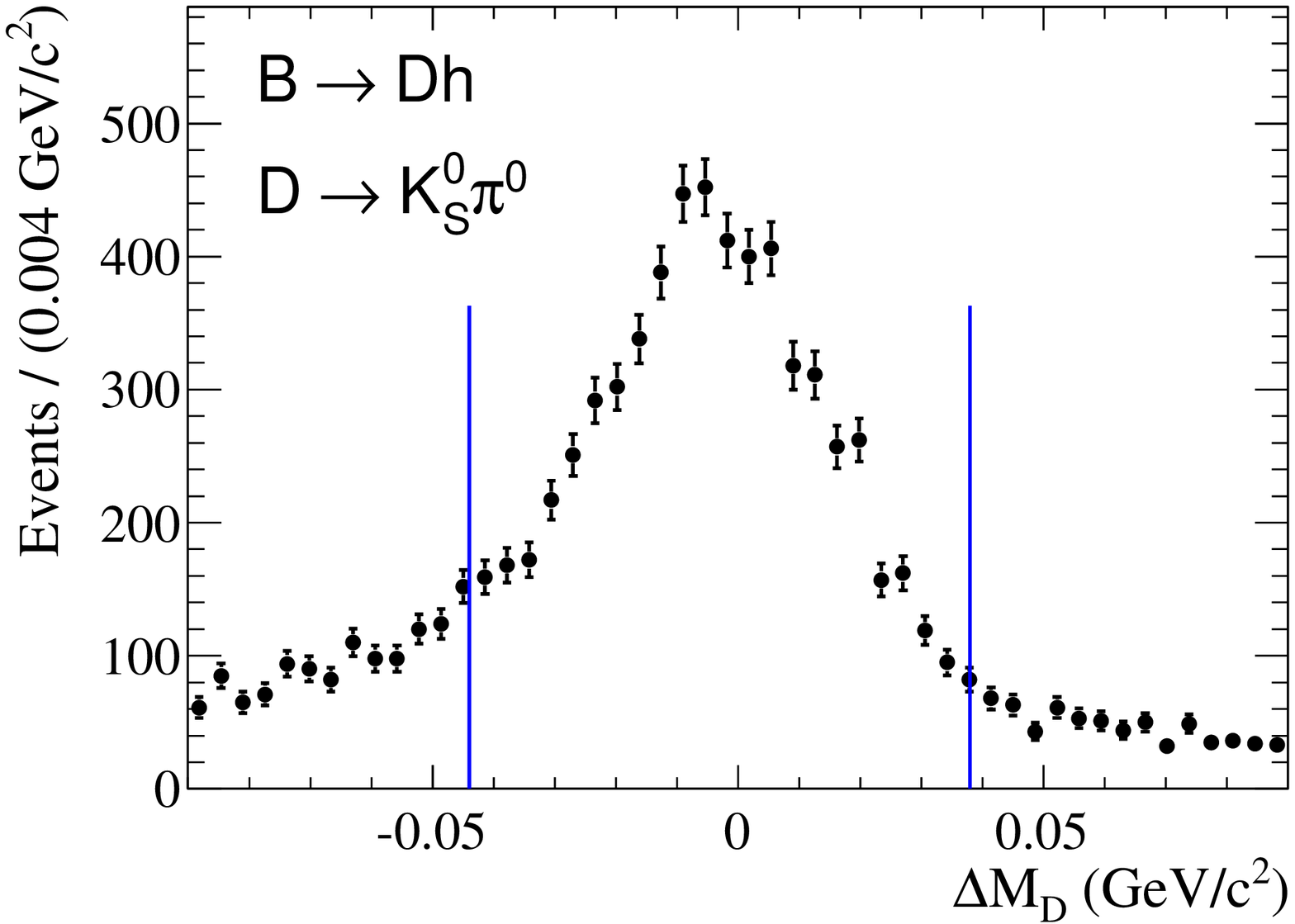}
\includegraphics[width=0.39\textwidth]{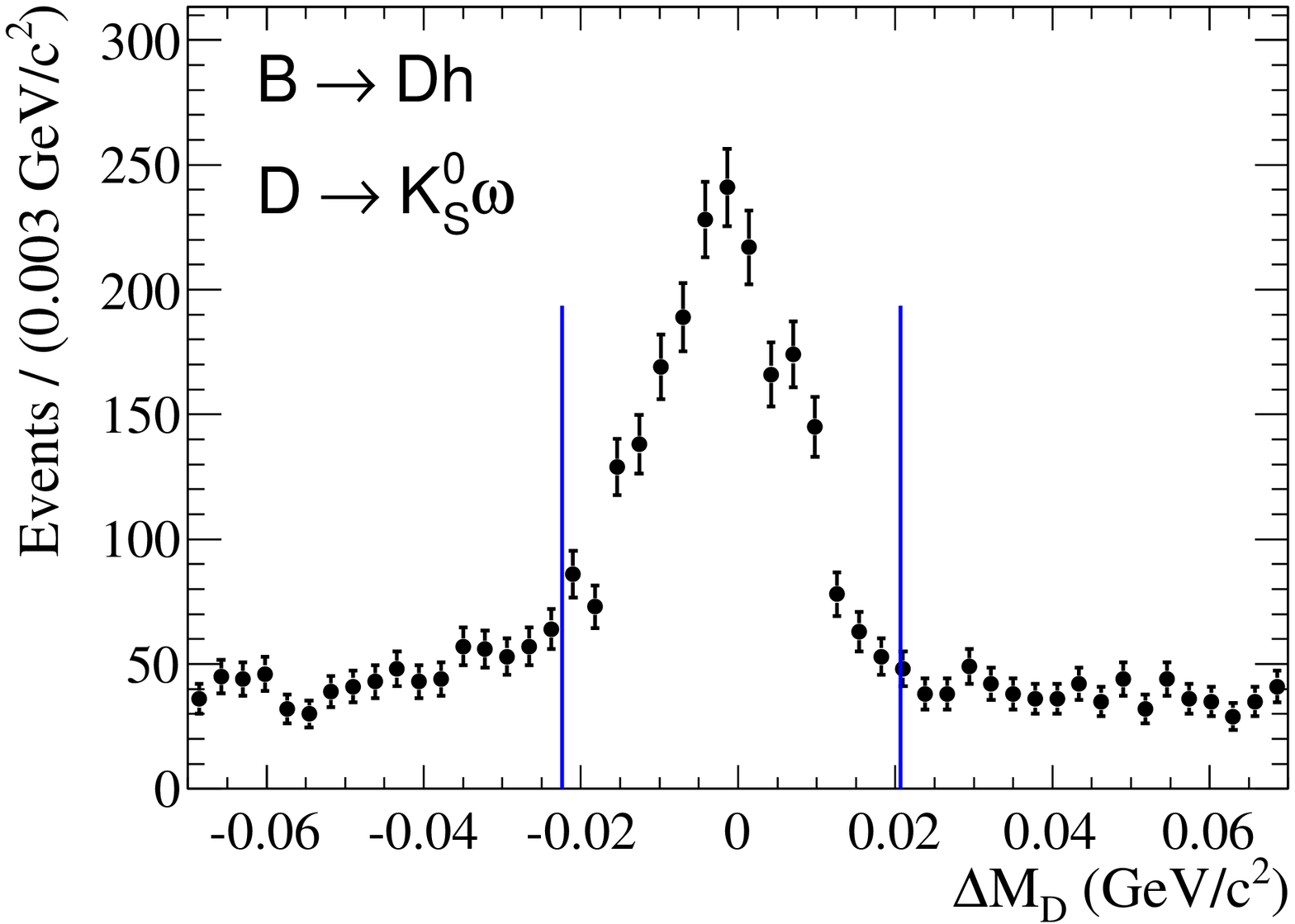}
\includegraphics[width=0.39\textwidth]{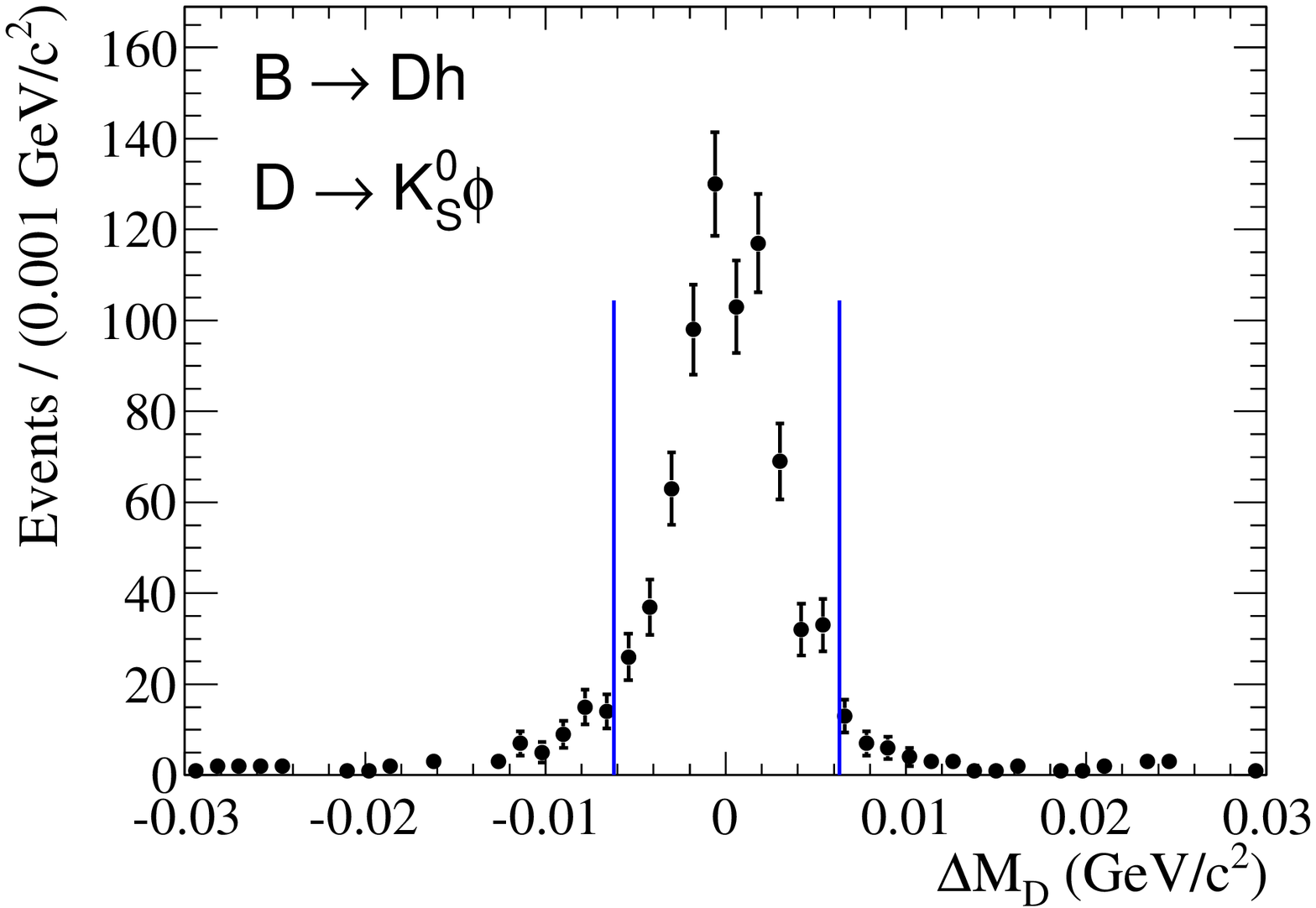}
\includegraphics[width=0.39\textwidth]{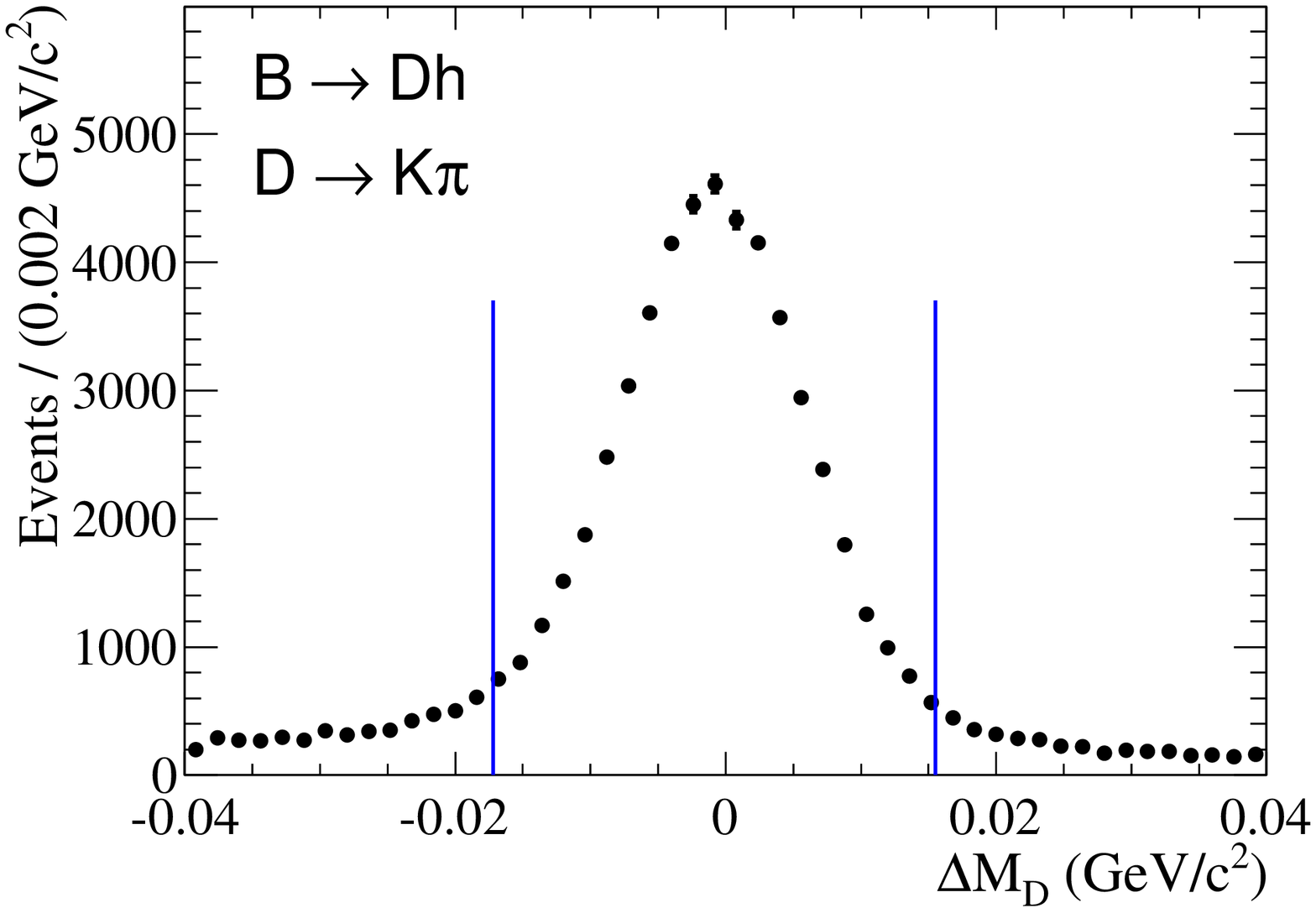}
\caption{Distributions of the difference between the $D$ candidate's
  invariant mass and the nominal $D^0$ mass~\cite{PDG2008}, as measured in
  the $B^\pm\to Dh^\pm$ samples. All selection criteria described in
  Section~\ref{sec:selection}, except that on the $D$ invariant
  mass $M_D$, have been applied, including the \chisq-based candidate
  selection. In addition we reduce the background
  by requiring the fit variables to satisfy
$\mes>5.27\gevcc$, $\de>-0.05\gev$, and $\fshr>-0.25$.
  The $\Delta M_D$ selection requirements are
  depicted by the vertical lines.}
\label{fig:dmasses}
\end{center}
\end{figure*}

The invariant mass distributions of the reconstructed $D$ candidates,
after all the other selection criteria described in this section have
been applied, are shown in Fig.~\ref{fig:dmasses}.

We reconstruct $B^\pm$ meson candidates by combining a neutral $D$ candidate
with a track $h^\pm$. For the $D{\to}K\pi$ mode, the charge of the track $h$
must match that of the kaon from the $D$ meson decay. This selects
$\b\to\c$ mediated $B$ decays $B^-\to \Dz h^-$ and $B^+ \to \Dzb
h^+$. The contamination from $b\to u$ mediated $B$ decays followed by
doubly-Cabibbo-suppressed $D$ decay, \textit{i.e.} 
$B^- \to \Dzb K^-$, $\Dzb \to K^-\pi^+$, and from \Dz--\Dzb mixing 
is negligible.
In the $B^\pm\to Dh^\pm$, $D\to\pipi$ channel we require that the
invariant mass of the $(h^\pm\pi^\mp)$ system is greater than 1.9\gevcc
to reject background from
$B^-{\to}\Dz\pi^-$, $\Dz{\to}K^-\pi^+$ and $B^-{\to}K^{*0}\pi^-$,
$K^{*0}{\to}K^-\pi^+$ decays and their \CP conjugates.
Here $\pi$ is the pion from the $D$ and $h$ is the track from the $B$
candidate taken with the kaon mass hypothesis.
To improve the $B$ momentum resolution, the neutral $D$
invariant mass is constrained to the nominal \Dz mass~\cite{PDG2008} for
all $D$ decay channels.

We identify signal $B\to DK$ and $B\to D\pi$ candidates using two kinematic
variables: the difference between the CM energy of the $B$ meson
($E^*_B$) and the beam energy,
\begin{equation}
	\Delta E=E^*_B - \sqrt{s}/2\ ,
\end{equation}
and the beam-energy-substituted mass,
\begin{equation}
	\mes = \sqrt{(s/2 + \mathbf{p}_{ee}\cdot\mathbf{p}_B)^2/E_{ee}^2-p_B^2},
\end{equation}
where $(E_B, \mathbf{p}_B)$ and $(E_{ee}, \mathbf{p}_{ee})$ are the 
four-momenta of the $B$ meson and of the initial \epem\ system, respectively, 
measured in the laboratory frame.
The \mes distributions for $B^\pm\to Dh^\pm$ signals are
centered at the $B$ mass~\cite{PDG2008}, have a root-mean-square
of approximately 2.6\mevcc, and do not depend strongly on either
the $D$ decay mode or the nature of the track $h$.
In contrast, the \de distributions depend on the
mass assigned to the track $h$. We evaluate \de with the kaon
mass hypothesis so that the peaks of the distributions are centered near
zero for $B^\pm\to DK^\pm$ events and are shifted by approximately $+50\mev$ 
for $B^\pm\to D\pi^\pm$ events. The \de resolution depends on the 
kinematics of the decay, and is typically
16\mev for all $D$ decay modes under study after the $D$ invariant
mass is constrained to its nominal value. We retain $B$ candidates
with $\mes$ and $\DeltaE$ within the intervals
$5.20<\mes<5.29\gevcc$ and $-80<\DeltaE<120\mev$, which define the
region for the fit described later.

In order to discriminate the signal from $e^+e^- \to q\bar{q}$
background events, denoted $q\bar{q}$ in the
following, we construct a Fisher discriminant \fshr
based on the four event-shape quantities $L_{20}^{\rm ROE}$,
$|\cos\theta_T^*|$, $|\cos\theta_B^*|$ and $H_{20}^{\rm ROE}$. 
These quantities, evaluated in the CM frame, are defined as:
\begin{itemize}
\item $L_{20}^{\rm ROE} = L_2/L_0$ is the ratio of the second and
  zeroth event shape moments of the energy flow in the rest of event
  (ROE), \textit{i.e.} considering all the charged tracks and neutral clusters
  in the event that are not used to reconstruct the $B$ candidate. They are
  defined as $L_2=\sum_i 
  p_i\cos^2\theta_i$ and $L_0=\sum_i p_i$, where $p_i$ are
  the momenta and $\theta_i$ the angles of the charged and
  neutral particles in the ROE, with respect to the thrust
  axis of the $B$ candidate's decay products. The thrust axis
  is defined as the direction that maximizes the sum of the
  longitudinal momenta of the particles used to define it;
\item $\theta_T^*$ is the angle between the thrust axis of the $B$
  candidate's decay products and the beam axis;
\item $\theta_B^*$ is the angle between the $B$ candidate momentum and
  the beam axis;
\item $H_{20}^{\rm ROE} = H_2/H_0$ is the ratio of the second and
  zeroth Fox-Wolfram moments $H_2$ and $H_0$~\cite{foxwolfram}, computed
  using charged tracks and photons in the ROE.
\end{itemize}
The quantity \fshr is a linear combination of the four aforementioned
event-shape variables:
\begin{equation}
	\fshr = c_1 L_{20}^{\rm ROE} + c_2 |\cos\theta_T^*| + c_3 |\cos\theta_B^*| + c_4 H_{20}^{\rm ROE}\,.
\end{equation}
The values of the coefficients $c_i$ are the ones which maximize the separation
between simulated signal events and a continuum background sample
provided by off-resonance data, taken
$\approx40\mev$ below the \Y4S resonance. The maximum likelihood fit 
described in Section~\ref{sec:fit} is restricted to events with \fshr
within the interval $-1.5<\fshr<1.5$, to remove poorly reconstructed
candidates.

For events with multiple $B^\pm\to Dh^\pm$ candidates (about 16\% of the
selected events), we choose the $B$ candidate with the smallest $\chi^2
= \sum_c (M_c-M_c^{\rm PDG})^2/(\sigma^2_{M_c}+\Gamma^2_{c})$ formed
from the measured and true masses, $M_c$ and $M_c^{\rm PDG}$, of all
the unstable particles $c$ produced in the $B$ decay tree ($D$, \piz,
\KS, $\phi$, $\omega$), scaled by the sum in quadrature of the resolution
$\sigma_{M_c}$ of the reconstructed mass and the  
intrinsic width $\Gamma_{c}$. From simulated signal events, we find
that this algorithm has a probability to select the correct candidate
between 98.2\% and 99.9\% depending on the $D$ decay mode.
We also find that the algorithm has negligible effect on the $M_D$ distributions.

%
%
We compare the distribution of each selection variable in data and
simulated events after the requirements on all other
variables have been applied. In order not to introduce biases that may
artificially enhance the signal yield, we perform a blind study by
explicitly removing, in this comparison, events consistent with the
$B^\pm\to DK^\pm$ signal, $i.e.$ those with $|\mes - m_B|<10~\mevcc$, 
$|\DeltaE|<40~\mev$, $\fshr>-0.8$ and track $h$ passing kaon identification criteria. 
We find excellent agreement between data and simulated events,
both for events consistent with the $B^\pm\to D\pi^\pm$ signal
($|\mes - m_B|<10~\mevcc$, $|\DeltaE - 50~\mev|<40~\mev$, $\fshr>-0.8$
and track $h$ failing the kaon identification criteria) and for
background-like events. 
We correct for small differences in the means and widths of the
distributions of the invariant masses of the unstable particles and of
\mes and \DeltaE both when applying to data the selection criteria obtained
from simulated events and in the final fit described in the next section. 

The total reconstruction efficiencies, based on simulated $B^\pm\to D K^\pm$
events, are summarized in the second column of Table~\ref{tab:efficiencies}.
\begin{table}[!h]
\begin{center}
\caption{Reconstruction efficiency for $B\to DK$ from simulated events. 
We also quote the efficiency and purity in a signal-enriched subsample 
(see text for details).}
\label{tab:efficiencies}
\begin{tabular}{lccc}
\hline\hline\\[-2.5ex]
\Dz mode	& Efficiency after  & Efficiency in   & Purity in \\
                & full selection    & signal-enriched & signal-enriched \\ 
                &                   & subsample       & subsample \\
\hline
\kpi	  	& 52\% & 22\% & 96\% \\
\kk	  	& 44\% & 18\% & 85\% \\
\pipi	  	& 38\% & 17\% & 68\% \\
\kspiz  	& 24\% & 10\% & 83\% \\
\ksphi	  	& 20\% &  9\% & 91\% \\
\ksomega 	& 10\% &  4\% & 71\% \\
\hline\hline
\end{tabular}
\end{center}
\end{table}
For the reasons explained in Section~\ref{sec:glw_introduction}, the
efficiencies are 40\% to 60\% higher than in our previous study of the
same decay channels~\cite{babar_d0k_GLW_PRD}.
The efficiencies obtained for $B^\pm\to D\pi^\pm$ events from the simulation
are statistically consistent with those for $B^\pm\to D K^\pm$, where the $D$
meson is reconstructed in the same final state.
For illustration purposes we define a signal-enriched sample for each $D$ decay mode, containing all
$B^\pm\to Dh^\pm$ candidates satisfying the criteria $-40<\de<100\mev$, 
$0.2<\fshr<1.5$, $5.275<\mes<5.285\gevcc$, and whose daughter track $h$
passes charged kaon identification criteria. The
typical kaon efficiency is $\approx77\%$ and the pion misidentification rate
is $\approx2\%$.
The reconstruction efficiencies and the expected purities for the
signal-enriched subsamples, determined on simulated data, are listed in
Table~\ref{tab:efficiencies}.

\section{Maximum Likelihood Fit}
\label{sec:fit}

We measure $R_{K/\pi}^{(\pm)}$
and \Acppm using simultaneous extended and unbinned maximum
likelihood fits to the distributions of the three variables \de, \mes,
and \fshr of $B$ candidates selected in data. The dataset is split into
24 subgroups by means of three discrete variables: the
charge $\eta=\pm 1$ of the reconstructed \B meson ($\times$2 subgroups);
the two-body $D$ decay final state $X$ ($\times 6$), allowing for a
more accurate description of the corresponding probability density
functions compared to the larger \CPpm subgroups;
and a PID variable
denoting whether or not the track $h$ from the $B$ passes ($p$) or fails ($f$)
charged kaon identification criteria ($\times 2$).
The pion misidentification rate of these criteria is determined
directly from data as described later, and is expected from
simulation to be around 2\%. The corresponding kaon identification
efficiency is $(77\pm 1)\%$, as determined from the signal MC
samples after weighting the bidimensional distribution of the
momentum and polar angle of the track $h$ by the ratio of the
analogous distributions observed in MC and data kaon control samples.
The uncertainty on the kaon identification efficiency is dominated by
the systematic contribution from the uncertainties on the weights.
We perform in total three simultaneous fits to these 24
subgroups: one fit for the two \CP-even $D$ final states (8
subgroups), one for the three \CP-odd $D$ final states (12 subgroups), and one
for the $D\to K\pi$ decay (4 subgroups).

The likelihood function $\mathcal{L}$ for each of these simultaneous
fits has the form
\begin{equation}
  \label{eq:fit_likelihood}
  \mathcal{L({\vec \nu})} = \frac{e^{-N}N^n}{n!} \prod_{s} \prod_{i=1}^{N_s} \mathcal{P}_s(m_{{\rm ES},i}, \de_i, \fshr_i; {\vec \nu}),
\end{equation}
where $s$ ranges over the subgroups under consideration, $N_s$ is the
number of events in subgroup $s$, $n$ is the total number of events in the
fit $n=\sum_s N_s$, and $N$ is the expected number of events. We
minimize $-\ln\mathcal{L}$ with respect to the set of fit parameters
${\vec \nu}$ specified later. The
probability $\mathcal{P}_{s,i} \equiv \mathcal{P}_s(\mes_i, \de_i,
\fshr_i)$ for an event $i$ is the sum of six signal and background
components: $B^\pm\to DK^\pm$ signal, $B^\pm\to D\pi^\pm$ signal,
background candidates from
$\epem\to\qq$ events, irreducible background arising from charmless
$B^\pm\to XK^\pm$ and $B^\pm\to X\pi^\pm$ decays, and background candidates
from other \BB events (reducible \BB background):  
\begin{eqnarray}
  \label{eq:pdf}
  N_s \mathcal{P}_{s,i} &=& N_s^{D\pi} \mathcal{P}_{s,i}^{D\pi} +
  N_s^{DK} \mathcal{P}_{s,i}^{DK} + \nonumber \\
  && N_s^{\qq}   \mathcal{P}_{s,i}^{\qq}   + N_s^{\BB}   \mathcal{P}_{s,i}^{\BB}   + \nonumber\\
  && N_s^{X\pi}  \mathcal{P}_{s,i}^{X\pi}  + N_s^{XK}
  \mathcal{P}_{s,i}^{XK}, 
\end{eqnarray}
where the $N_s^j$ are the expected yields in each
component $j$. In case of negligible correlations among the fit
variables, each probability density function (PDF) $\mathcal{P}$ factorizes as:
\begin{equation}
  \label{eq:pdffactorize}
  \mathcal{P}(\mes, \de, \fshr) = \mathcal{P}(\mes) \, \mathcal{P}(\de) \, \mathcal{P}(\fshr).
\end{equation}

The irreducible \BB background originates from events where a $B$ meson
decays to the same final state $Xh$ as the signal, but without the
production of an intermediate charmed meson in the decay chain.
When exploiting the \de, \mes, and \fshr variables, this background is
therefore indistinguishable from the signal.
As an example, the decay $B^\pm\to \Kp\Km K^\pm$ ($X=\kk$) is
an irreducible background for $B^\pm\to D_{\CP+}K^\pm$, $D_{\CP+}\to\kk$.
As described later in Section~\ref{sec:peaking_bkg}, the irreducible
background yield can be estimated by studying sideband regions of the $D$ candidate 
invariant mass distribution, and can then be fixed in the final fit,
where we assume $\mathcal{P}_i^{Dh} = \mathcal{P}_i^{Xh}$.

We express the signal yield parameters $N^{DK}_s$ and $N^{D\pi}_s$
through the \CP asymmetries $A^X_{DK}$ and $A^X_{D\pi}$ of $B^\pm\to DK^\pm$, 
$D{\to}X$ and $B^\pm\to D\pi^\pm$, $D{\to}X$, their branching fraction ratios, 
$R_{K/\pi}^X$, the total number $N^{D\pi}_{{\rm tot},X}$ of 
$B^\pm\to D\pi^\pm$, $D\to X$ signal events, the true kaon
identification efficiency $\varepsilon$ of the PID selector, and the
pion misidentification rate $m$ of the PID selector: 
\begin{eqnarray}
	\label{eq:yields1}
	N^{DK}_{\eta,p,X} &=& \frac{1}{2} \left(1 - \eta A^{X}_{DK}\right) \: N^{D\pi}_{{\rm tot},X} \: \Rkp^X \, \varepsilon\,,\\
	\label{eq:yields2}
	N^{DK}_{\eta,f,X} &=& \frac{1}{2} \left(1 - \eta A^{X}_{DK}\right) \: N^{D\pi}_{{\rm tot},X} \: \Rkp^X \, (1-\varepsilon)\,,\\
	\label{eq:yields3}
	N^{D\pi}_{\eta,p,X} &=& \frac{1}{2} \left(1 - \eta A^{X}_{D\pi}\right) \: N^{D\pi}_{{\rm tot},X} \: m\,,\\
	\label{eq:yields4}
	N^{D\pi}_{\eta,f,X} &=& \frac{1}{2} \left(1 - \eta A^{X}_{D\pi}\right) \: N^{D\pi}_{{\rm tot},X} \: (1-m)\,.
\end{eqnarray}
Because the ratios $R_{K/\pi}^X$ are small, the fit is not able to
determine the value of $\varepsilon$. Therefore we fix it to the aforementioned
value of $\varepsilon=(77\pm 1)\%$.
The reconstruction and selection efficiencies for true $B^\pm\to DK^\pm$ and
$B^\pm\to D\pi^\pm$ candidates, where the $D$ meson decays to the same final
state, are assumed to be identical. A systematic uncertainty is
assigned due to this assumption (see Section~\ref{sec:systematics}). The
simultaneous fit to the two \CP-even modes constrains
\begin{eqnarray}
	A^{\pipi}_{DK} &=& A^{\kk}_{DK} \equiv A_{\CPp}\,,\\
	\Rkp^{\pipi} &=& \Rkp^{\kk} \equiv \Rkp^+\,,
\end{eqnarray}
while the simultaneous fit to the three \CP-odd modes constrains
\begin{eqnarray}
	A^{\kspiz}_{DK} &=& A^{\ksphi}_{DK} = A^{\ksomega}_{DK} \equiv A_{\CPm}\,,\\
	\Rkp^{\kspiz} &=& \Rkp^{\ksphi} = \Rkp^{\ksomega} \equiv \Rkp^-\,.
\end{eqnarray}

The \mes distributions of the signal components are parameterized
using an asymmetric Gaussian shape, \textit{i.e.} a Gaussian with different
widths on both sides of the peak.
We use the same shape for $B^\pm\to DK^\pm$ and $B^\pm\to D\pi^\pm$, 
so the \mes $B^\pm\to DK^\pm$ signal shape (whose parameters are
floating in the fit) will mostly be determined by the much more
abundant $B^\pm \to D\pi^\pm$ control sample. Since the 
selection efficiencies for the two channels are the same, we expect the
number of reconstructed candidates from $B^\pm \to D\pi^\pm$ to be about twelve
times higher than for $B^\pm \to DK^\pm$.
We have checked that the \mes shapes for $B^\pm\to DK^\pm$ and $B^\pm\to
D\pi^\pm$ are consistent, and that the assumption that they are
identical does not bias the parameters of interest.

The \de distribution of the $B^\pm\to DK^\pm$ signal component is
parameterized with a double Gaussian shape.
The core Gaussian has a mean close to zero, a width
around 16\mev and, according to the simulation, accounts for about
90\% of the true $B^\pm\to DK^\pm$ candidates. The second Gaussian accounts for
the remaining 10\% of candidates whose energy has been poorly
measured.
The mean and
width of the core Gaussian are directly determined from data, while
the remaining
three parameters (the difference between the two means, the ratio
between the two widths and the ratio of the integrals of the two
Gaussian functions) are fixed from the simulation. In contrast to the
\mes case, the $B^\pm\to D\pi^\pm$ \de shape is not the same as for
$B^\pm\to DK^\pm$. This is due to the fact that we always assign the kaon
mass hypothesis to the track: the wrong mass assignment, in the
case of $B^\pm\to D\pi^\pm$, 
introduces a shift to the reconstructed energy of the pion and thus to
\de, since $\de = E^*_B - \sqrt{s}/2 = E^*_D + E^*_h -\sqrt{s}/2$. The
shift depends on the magnitude of the momentum ${\bf p}$ of the 
track $h$ in the laboratory frame, 
\begin{equation}
  \label{eq:deshift}
  \deshift({\bf p})=\gamma_{\textrm{\small CM}}
  \left(\sqrt{m_K^2+{\bf p}^2} -
  \sqrt{m_{\pi}^2+{\bf p}^2}\right).
\end{equation}
Therefore we parameterize the $B^\pm\to D\pi^\pm$ \de signal component 
with the sum of two Gaussians whose means are computed event-per-event 
by adding $\deshift({\bf p})$ to the means of the Gaussian functions used to
describe the $B^\pm\to DK^\pm$ \de signal.
The other parameters of the $B^\pm\to D\pi^\pm$ and $B^\pm\to DK^\pm$
\de distributions (the two widths and the ratio of the
integrals) are identical. Again, we exploit the $B^\pm\to D\pi^\pm$ control
sample to determine the shape of the $B^\pm\to DK^\pm$ signal.
In the case of the high
statistics flavor mode $D\to K\pi$, we add a linear background component to the double
Gaussian shape to account for misreconstructed
events, which peak in \mes but not in \de. The ratio between the
integral of the linear component and that of the two Gaussian
functions is fixed from simulated signal events. 

For the reducible \BB background, Eq.~\ref{eq:pdffactorize} does not hold
because of significant correlations between the \de and \mes distributions.
This reflects the fact that this background is composed of two
categories of $B$ candidates with different \mes and \de distribution:
\begin{itemize}
\item $B$ candidates formed from random combinations of charged
  tracks and neutral objects in the event, which populate the whole
  \mes-\de plane;
\item $B$ candidates from $B^\pm\to D\rho^\pm$, $B^\pm\to DK^{*\pm}$,
  $B^\pm\to D^{*}h^\pm$ $(D^*\to D\pi)$, where a pion from the $\rho$, 
  $K^*$ or $D^*$ decay is not reconstructed.
  These candidates peak in \mes close to the $B$ mass, but
  with broader resolution compared to the signal, and are shifted
  towards negative \de values, typically peaking at $\de \approx
  -m_\pi c^2$, therefore outside of the \de fit region; however, the tail
  on the positive side of the distribution extends into the \de fit region.
\end{itemize}
We parametrize the \mes-\de distribution of the \BB background by
means of two factorizing components: 
\begin{eqnarray}
  \mathcal{P}_{\BB}(\mes, \de) = f &\times& g_{\rm peak}(\mes) h_{\rm peak}(\de)+ \nonumber\\
  (1-f) &\times& g_{\rm cont}(\mes) h_{\rm cont}(\de).
\end{eqnarray}
The \mes component of the peaking part, $g_{\rm peak}(\mes)$, is
parameterized with a Gaussian function for $X=\pipi,\ \ksomega,\ \ksphi$.
For $X=\kk,\ \kspiz$ we use the ``Crystal Ball''
lineshape~\cite{crystalball}, an empirical smooth function that better
describes the non-Gaussian tail on the negative side of the distribution,
\begin{equation}
	\label{eq:cb}
	C(x) = \left\{ \begin{array}{ll}
	\frac{n^n}{|\alpha|^n} e^{-\frac{|\alpha|^2}{2}} \left(\frac{n}{|\alpha|}-|\alpha|-\bar{x} \right)^{-n}
		& \bar{x}<-|\alpha|\,,\\
	\exp\left(-\frac{1}{2}\bar{x}^2\right)\,
		& \bar{x}\ge-|\alpha|\,,
	\end{array} \right.
\end{equation}
with $\bar{x}=(x-\mu)/\sigma$ and $\bar{x}\to-\bar{x}$ for $\alpha<0$.
For $X=K\pi$ we use an empirical function of the form:
\begin{equation}
	N(x) = \exp\left(-\frac{1}{2\tau^2} \, \left\{ \ln^2[1 + \Lambda \tau (x - \mu)] + \tau^4 \right\} \right)\,,
\end{equation}
with $\Lambda = \sinh(\tau\sqrt{\ln 4})/(\sigma\tau\sqrt{\ln 4})$. Here 
$\mu$ is the position of the peak, while $\sigma$ and $\tau$ are parameters 
related to the width of the distribution on the two sides of the peak.
The
\de component of the peaking part $h_{\rm peak}(\de)$ is described with
a simple exponential function for the five \CP self-conjugate $D$
final states, and with a Landau function for the non-\CP-eigenstate final
state. The \BB purely combinatorial background component is described by
the 2-dimensional product of a linear background, $h_{\rm cont}(\de)$, and an 
empirical
function introduced by the ARGUS collaboration~\cite{argus}, $g_{\rm
cont}(\mes)=A(\mes/m_0)$:
\begin{equation}
	A(x) = x (1-x^2)^p \exp\left(-\zeta\left[ 1 - x^2\right]\right)\ ,
\end{equation}
where $m_0 = \sqrt{s}/(2c^2) = 5.29\gevcc$ is the kinematic
endpoint of the \mes distribution. All the parameters of the \BB
background \mes-\de distribution are fixed from simulated \BB events.
The only exception is the width of the Landau function used for $h_{\rm
peak}(\de)$ in $X=\kpi$. This parameter controls the behaviour at low
\de values, $\de\approx -80\mev$, where we find the simulation not to be
sufficiently precise given the high statistics of this channel. We note
that the shape parameters differ across the six final states, but are
similar across the charge and PID selector subgroups belonging to one
final state.

In \qqbar events, $B$ candidates arise from random combinations of
charged tracks and neutral particles produced in the hadronization of
the light quark-antiquark pairs produced in $e^+e^-$ collisions.
Similarly to the combinatorial component of the \BB background, the
\qqbar background distribution in the \mes-\de plane is parameterized by
the product of an ARGUS function in \mes and a linear background in \de. We
float the slope of the linear components, while the parameters of the ARGUS
function are fixed, in each $D$ final state, from simulated \qqbar
events. They are in good agreement across the final states and other subgroups.

The \fshr distributions are parameterized in a similar way for all fit
components. We find that the distributions of $B^\pm \to DK^\pm$ and
$B^\pm\to D\pi^\pm$ signal events are consistent with each other, as 
expected since their kinematics are very similar, and choose 
to parameterize them with the same shape. 
For this we use the sum of two asymmetric Gaussian functions. Some channels
with lower statistics don't require the full complexity of this
parameterization: in those cases we use a single asymmetric
Gaussian, a double Gaussian, or a single Gaussian.
In particular we use:
for the signal components a double asymmetric Gaussian, except for
$X=\ksphi$, where a double Gaussian function is adopted;
for the \BB background components a double asymmetric Gaussian in
case of $X=\kspiz,\kpi$, a double Gaussian in case of $X=\ksomega$, and
a single Gaussian otherwise;
for the \qq background components a double asymmetric Gaussian,
except for $X=\ksphi$, where we use a single Gaussian.
%

In summary, the floating parameters of the fits are:
all parameters related to the signal yields, and therefore to the
GLW parameters, as given in Eqns.~\ref{eq:yields1}-\ref{eq:yields4}, except
$\varepsilon$; 
all background yields and \CP-asymmetries except the irreducible
background yields and asymmetries, the \BB asymmetries for \CPm modes
and for the $(\CPp, p)$ subgroups ($B\to D_{\CP +}h$ candidates where the
track $h$ passes the kaon identification criteria),
and the \BB yield in the $(\ksphi, p)$ subgroup;
selected shape parameters, namely the overall width and mean of
the \de signal, the \mes signal shape, and the \de and \fshr shape for
\qq background.
A full list of the floating parameters can be found in
Table~\ref{tab:fitresult}.
The non-floating parameters are fixed to their expectations obtained from
simulation or, in case of the irreducible background yields, to values
obtained from data control samples (see next section).
Non-floating \CP asymmetries are fixed to zero. We assign systematic
uncertainties due to the fixed parameters. 

We check that the fitter is correctly implemented by generating and fitting a
large number of test datasets using the final PDFs. In this study, we
include an analytic description for the conditional variable \deshift.
The residuals for a given parameter, divided by the measured parameter
error, should follow a Gaussian distribution with zero mean ($\mu$) and
unitary width ($\sigma$). We observe no significant deviations from the expected distribution.
In particular, \Rkpp shows the largest shift from zero mean
($\mu = -0.06 \pm 0.07$) and \Acpp shows the largest deviation from
unity width ($\sigma = 1.13 \pm 0.06$) among the parameters of interest.

We investigate fit biases, arising from possible
discrepancies between the true signal distribution and the chosen fit model,
by fitting a large number of test datasets, in
which the $B^\pm\to DK^\pm$ and $B^\pm\to D\pi^\pm$ signal components 
are taken from simulated samples of sufficient statistics, while the background
components are randomly generated according to their PDFs. 
Of all floating parameters,  
only \Rkp acquires a significant bias, resulting in corrections of 0.5
and 1.0 times the expected statistical uncertainties on these
parameters in the \CP and flavor modes, respectively.
This bias is caused by small differences
in the \de distributions of the signal components across the kaon PID subgroups ($p$ and $f$),
which the final PDF does not account for. A second, smaller contribution
to this bias is a small discrepancy between the \dep signal shape of
$B^\pm\to D\pi^\pm$ events and the \dek shape of $B^\pm\to DK^\pm$ events.
The biases in the \Rkp parameters are correlated, and partly cancel in
the ratio, resulting in a smaller bias on the GLW parameters \Rcppm.
The largest (smallest) remaining bias is 0.12 (0.05) times the expected
statistical uncertainty for \Rcpp (\Acpm).
We correct the final values
of the parameters \Acp and \Rkp for the observed biases, and assign 
systematic uncertainties to these corrections.

\section{Irreducible background determination}
\label{sec:peaking_bkg}
As discussed in the previous section, the irreducible background
arises from charmless $B^\pm\to X h^\pm$ decays, which have the same final
states as the $B^\pm\to D(\to X) h^\pm$ signal and therefore the same
distribution of the three fit variables \de, \mes, and \fshr.

In the \dztokpi flavor mode, the irreducible background -- taking into
account the measured branching fractions for $B^\pm\to K^\pm\pi^\mp K^\pm$ and
$B^\pm\to K^\pm\pi^\mp\pi^\pm$~\cite{PDG2008} and a selection efficiency of
$\approx 1\%$, estimated from simulated events -- is 
negligible compared to the expected signal yields (about 3400 $B^\pm\to DK^\pm$
and 45000 $B^\pm\to D\pi^\pm$ expected signal events).
On the other hand, in the \CP modes, where the signal yields are
expected to be an order of magnitude lower than in \kpi, and the upper limits
for the branching ratios of $B^\pm\to X h^\pm$ decays are at the $10^{-5}$ level, we cannot \textit{a priori}
exclude a relevant irreducible background contribution. 

We estimate the irreducible background yields in our sample by
exploiting the fact that the $D$ invariant mass distribution for this
background is approximately uniform, while for the signal it is
peaked around the nominal $D$ mass. Therefore we can select a control
sample containing irreducible background candidates, but with the signal
strongly suppressed, by applying the same selection as for the signal,
with the only difference that the $D$ invariant mass is required to lie
in a region ($D$ invariant mass sidebands) which is separated by at least a
few $\sigma_{M_D}$ from the nominal $D$ mass
(see Table~\ref{tab:sideband}). 
We then perform an extended maximum
likelihood fit to the \mes, \de, and \fshr distributions of the control
sample in order to measure the irreducible background yields in the $D$
invariant mass sidebands. 
The fit is similar to the nominal one described in the previous section.
However, due to the limited statistics available in the sidebands, we
are forced to fix more parameters compared to the nominal fit; in
particular, we fix any possible charge asymmetry of the $B^\pm\to Xh^\pm$
decays to zero (a systematic uncertainty is assigned to this
assumption).
Finally, since the $D$ candidate invariant mass distribution of the irreducible
background is approximately uniform, we scale the obtained yields by
the ratio of the widths of the $D$ signal and control sideband mass
regions to obtain the irreducible background yield $N^{Xh}$ (scale
factor in Table~\ref{tab:sideband}).
Table~\ref{tab:pkbg_result} shows the scaled irreducible
background yields that enter the final fit.

\begin{table}[!ht]
\center
\caption{$D$ mass sideband definitions, the scale factor defined
as the ratio of the widths of the $D$ mass signal and sideband
regions.} 
\label{tab:sideband}
\begin{tabular}{lcc}
\hline\hline
$D$ decay& $M_D$ sideband region                & Scale   \\
mode     & (\mevcc)                             & factor  \\
\hline
\kk      & $[1794.5-1834.5]$, $[1884.5-1914.5]$ & 0.43    \\
\pipi    & $[1814.5-1839.5]$, $[1889.5-1934.5]$ & 0.48    \\
\kspiz   & $[1774.5-1804.5]$, $[1924.5-1954.5]$ & 1.67    \\
\ksomega & $[1794.5-1829.5]$, $[1899.5-1934.5]$ & 0.69    \\
\ksphi   & $[1794.5-1834.5]$, $[1894.5-1934.5]$ & 0.28    \\
\hline\hline
\end{tabular}
\end{table}

\begin{table}[!ht]
\begin{center}
\caption{Irreducible background yields estimated from $M_D$ sidebands in
data.}
\label{tab:pkbg_result}
\begin{tabular}{lr@{~$\pm$~}lr@{~$\pm$~}l}
\hline\hline\\[-2.5ex]
$D$ decay mode	& \multicolumn{2}{c}{$N^{XK}$} & \multicolumn{2}{c}{$N^{X\pi}$} \\
\hline
\kk	  	&   93  & 10   	& $-5$  &  8   	  \\
\pipi	  	&    4  &  6   	&   0   &  9   	  \\
\kspiz	  	& $-4$  &  9   	&  65   & 23   	  \\
\ksomega  	&    3  &  6   	&   0   &  8   	  \\
\ksphi	  	&    0.5&  0.7 	&   1.4 &  1.0 	  \\
\hline\hline
\end{tabular}
\end{center}
\end{table}

\section{Systematic Uncertainties}
\label{sec:systematics}

We consider nine sources of systematic uncertainty that may affect
the GLW parameters \Acppm and \Rcppm. Their contributions are summarized
in Table~\ref{tab:syst_summary}.

First, we estimate the influence of fixed parameters of the nominal
PDF. We perform a large number of test fits to the
data, similar to the nominal fit. In each of these test fits the fixed
parameters are varied according to their covariance matrices. From the
resulting distributions we calculate the systematic covariances of the fit
parameters \Acppm and \Rkp. The parameters responsible for the largest
uncertainty are the \mes endpoint $m_0$, and parameters related to the
measured yields, \emph{e.g.} \bb background asymmetries and the efficiency of the
kaon selector.

The uncertainties in the irreducible background event yields introduce a
systematic uncertainty in the $B^\pm\to D_{\CP}h^\pm$ yields and therefore 
in \Rcppm.
Likewise, any charge asymmetry in this background would
affect the measured values of \Acppm. We again perform a series of test
fits to on-peak data, where we vary the $B^\pm\to Xh^\pm$ yields and 
asymmetries by their uncertainties. For the latter, we take the uncertainties
to be $\pm 10\%$ for $X=\kk$ and $\pm 20\%$ for the other modes, which are
conservative estimates consistent with the existing upper limits on the 
\CP asymmetries in those decays~\cite{HFAG}.
For \Acpp, the possible \CP asymmetries in the peaking background
dominate the systematic error.

As explained in Section~\ref{sec:fit}, we correct the fit results for
biases observed in Monte Carlo studies. We take the associated  
systematic uncertainties to be half the size of the bias corrections,
summed in quadrature with the statistical uncertainties on the biases.
The latter are due to the limited number of test fits used to estimate
the corrections.

We investigate a potential charge asymmetry of the \babar\ detector, due
to a possible charge bias in tracking efficiency (e.g. \Kp vs \Km)
and/or particle identification. Our analysis includes a number of
control samples, in which the \CP asymmetry is expected to be negligible:
the six $B^\pm\to D\pi^\pm$ samples and the $B^\pm \to DK^\pm$ flavor mode 
($D\to K\pi$). 
The weighted average of the charge asymmetry in the control samples
is $(-0.95 \pm 0.44)\%$, from which we assign uncertainties of 1.4\%
to both \Acpp and \Acpm.
We consider these uncertainties to be 100\% correlated.

The measured \CP asymmetry in $B^\pm\to DK^\pm$, $D\to\ksphi$, can be diluted 
by the presence of $B^\pm\to DK^\pm$ decays followed by $D$ decays to the same
final state $\KS\kk$ as the signal but with opposite \CP content,
such as $D\to\KS a_0$, $a_0\to\kk$. The
same can happen in the $B^\pm\to DK^\pm$, $D\to\ksomega$ analysis with
backgrounds from $B^\pm\to DK^\pm$, $D\to\KS\pipipiz$. 
This background can also affect the ratios \Rcpm. It is
possible to obtain correction factors to both \Acpm and \Rcpm from a fit
to the distributions of the relevant helicity angles, $\cos\theta_N$ and
$\cos\theta_H$ for \ksomega and \ksphi, respectively. The fit is
performed on dedicated $B^\pm\to D\pi^\pm$ samples, in which the selection
requirements on the helicity angles have not been applied. It can be
shown~\cite{tesiGM} that for these two final states the observed
charge asymmetries and ratios should be corrected by a factor
\begin{eqnarray}
	\label{eq:acpcorr}
	\Acp^{\rm true} &=& \Acp^{\rm obs} \cdot \frac{1 + f_{\epsilon}|z|^2 R'}{1 - f_{\epsilon}|z|^2},\\
	\label{eq:rcpcorr}
	\Rkp^{\rm true} &=& \Rkp^{\rm obs} \cdot \frac{1 + f_{\epsilon}|z|^2}{1 + f_{\epsilon}|z|^2 R'}.
\end{eqnarray}
Here, $R'$ is the ratio of the \Rkppm values, where \Rkpm is taken from a
single fit to the \dztokspiz final state only (as opposed to using all
three \CPm final states under study), $R' = \Rkpp/\Rkp^{\kspiz}$, and
$f_{\epsilon}=\epsilon_{\rm sig}/\epsilon_{\rm bkg}$ is the ratio of the
efficiencies of the selection criterion on the helicity angles:
$f_{\epsilon,\ksomega}=0.71$ and $f_{\epsilon,\ksphi}=0.64$.
To apply
these corrections, we first perform a fit of the \kspiz final state
alone to obtain $\Rkp^{\kspiz}$. We then perform the simultaneous fit of
the \CPp final states, from which we take the value of \Rkpp. Finally, we
include the correction factors into the \CPm final PDF, which will allow
the likelihood fitter to correctly estimate their influence. The
parameter $|z|^2$ in Eqns.~\ref{eq:acpcorr} and~\ref{eq:rcpcorr} is
extracted from fits of the helicity angle distributions in the
\dztoksomega and \dztoksphi subsamples to the function
$|z|^2+3\cos^2\theta$~\cite{tesiGM}.
We subtract the 
background expected from the Monte Carlo simulation, which
has been rescaled to match the data. We find $|z|^2 = 0.065 \pm 0.033$ in
the case of \ksomega, and $|z|^2 = 0.217 \pm 0.063$ in the case of \ksphi. The
uncertainties contain propagated uncertainties due to the background
subtraction. The resulting corrections are:
\begin{eqnarray}
  A_{\CP(\ksomega)}^{\rm true}   & = & A_{\CP(\ksomega)}^{\rm obs}   \times (1.105  \pm 0.056)\,,\\
  A_{\CP(\ksphi)}^{\rm true}     & = & A_{\CP(\ksphi)}^{\rm obs}     \times (1.35   \pm 0.12)\,,\\
  R_{K/\pi(\ksomega)}^{\rm true} & = & R_{K/\pi(\ksomega)}^{\rm obs} \times (0.9929 \pm 0.0066)\,,\\
  R_{K/\pi(\ksphi)}^{\rm true}   & = & R_{K/\pi(\ksphi)}^{\rm obs}   \times (0.981  \pm 0.016)\,.
\end{eqnarray}
In order to assign systematic uncertainties, we propagate the uncertainties on the
correction factors into the final result.

When calculating \Rcp through Eq.~\ref{eq:rpmdef} one has to take
into account that this equation is an approximation.
We define the double ratios used to approximate $R_{\CP\pm}$ as
$R_{\pm}$. They are given by
\begin{eqnarray}
  \label{eq:syst_rpm}
  R_\pm&{=}&\frac{\Gamma(B^- \to D_{\CPpm}K^-)   + \Gamma(B^+ \to D_{\CPpm}K^+)}
                 {\Gamma(B^- \to D_{f} K^-)      + \Gamma(B^+ \to \overline{D}_{f} K^+)} \nonumber\\
  &{\times}&\frac{\Gamma(B^- \to D_{f} \pi^-)    + \Gamma(B^+ \to \overline{D}_{f} \pi^+)}
                 {\Gamma(B^- \to D_{\CPpm}\pi^-) + \Gamma(B^+ \to D_{\CPpm}\pi^+)},
\end{eqnarray}
where $D_f$ denotes the $K^-\pi^+$ final state.
These can be written as
\begin{eqnarray}
  \label{eq:syst_rpm2}
  R_\pm & = & \frac{1+r_B^2\pm2r_B\cos\delta_B\cos\gamma}{1+r_B^2r_D^2+2r_Br_D\cos(\delta_B{-}\delta_D)\cos\gamma} \nonumber\\
  & \times &\frac{1+r_{B\pi}^2r_D^2+2r_{B\pi}r_D\cos(\delta_{B\pi}{-}\delta_D)\cos\gamma}
                 {1+r_{B\pi}^2\pm2r_{B\pi}\cos\delta_{B\pi}\cos\gamma},
\end{eqnarray}
where $r_{B\pi}$ and $\delta_{B\pi}$ are defined, in analogy to $r_B$
and $\delta_B$, as
$r_{B\pi}e^{i(\delta_{B\pi}-\gamma)}=A(\Bm\to\Dzb\pim)/A(\Bm\to\Dz\pim)$,
while $r_D$ and $\delta_D$ are defined as $r_De^{i\delta_D} =
A(\Dzb\to\Km\pip)/A(\Dz\to\Km\pip)$. 
We write Eq.~\ref{eq:syst_rpm2} in the form $R_\pm =
\Rcppm \times (1+R_c)$, and we assign a relative systematic uncertainty
based on the value of the correction $R_c$. Taking $\sin\theta_C = 0.2257
\pm 0.0010$ (where $\theta_C$ is the Cabibbo angle) and
$r_B{=}0.104^{+0.015}_{-0.025}$ from~\cite{CKMfitter},
and expressing $r_D=|V_{cd}V_{us}|/|V_{ud}V_{cs}|=\tan^2\theta_C$,
and $r_{B\pi} = r_B \tan^2\theta_C$,
we find $R_c \approx 4 r_B \tan^2\theta_C
\approx 2.2\%$. Here, we have conservatively assumed values for the cosine
terms which maximize $R_c$.
We thus assign a relative uncertainty of $2.2\%$ to
the values of \Rcp, fully correlated between \Rcpp and \Rcpm.

We also consider the influence on the measured value of \Acp of 
misreconstructed signal $B$ candidates, $i.e.$ candidates reconstructed, 
in events containing a true $B\to DK$ decay with $D$ decaying to the same
final state $X$ as the reconstructed candidate, 
from random combinations of particles produced in the true $B\to DK$ decay
and the particles of the ROE.
The fraction of these candidates ranges from 0.3\% to 12\% in
simulated $B^\pm\to D_{\CP}K^\pm$ events, depending on the channel. 
Since we treat this component as signal, we implicitly assume
that its charge asymmetry is equal to the asymmetry in the signal
component.
We use simulated signal events to estimate the ratio between
misreconstructed and true $B^+\to DK^+$ candidates and the ratio between
misreconstructed and true $B^-\to DK^-$ candidates, and find these
two quantities to differ by less than 0.1\%, from which we derive an
upper limit on the difference between the observed and the true value
of \Acp.

The yield double ratios \Rcppm should be corrected by the corresponding
double ratio of selection efficiencies. We find from simulated events 
that the efficiency double ratios are compatible with each other, and their 
average value is 
very close to unity, $(99.46\pm0.23)\%$. Thus we do not correct the
central values but conservatively assign a relative uncertainty equal to
$1 - (0.9946 - 0.0023) = 0.0077$.

The final PDF doesn't contain an explicit description of the conditional
parameter \deshift, assuming implicitly that the distribution of
\deshift observed in data is the same for all the components of the fit.
However, the distributions are found to be slightly 
different across the components, thus introducing a possible bias in the
fit results.
To estimate the size of this bias, we use simulated events to obtain
parameterizations of the \deshift distributions of all the fit
components and repeat the fits to data. We assign the differences compared to the
results of the nominal fits as the systematic uncertainty.
We expect this effect to be highly correlated between \Acp
parameters, because the PDFs
are similar in each $D$ decay channel. Thus they are affected by non-uniform
\deshift distributions in a similar way. The same argument holds for the
\Rkp parameters. We studied the effect of assigning a 0\%, 50\%, and
100\% correlation. The uncorrelated case gave the largest deviations
from the nominal results, the fully correlated case gave the smallest.
However, the variation was found to be at the 10\% level. 
We assign the systematic uncertainty corresponding to a correlation of 50\%.

Table~\ref{tab:syst_summary} lists the contributions of the effects
discussed above. Compared to our previous
analysis~\cite{babar_d0k_GLW_PRD}, the systematic uncertainty on
\Acpp is reduced due to better understanding of the detector
intrinsic charge asymmetry (the determination of which benefits from
the larger dataset) and due to improved evaluation of the correlations
among the different sources of systematic uncertainties. The uncertainty
on \Acpm is only slightly reduced. By contrast, the
systematic uncertainties on \Rcppm are increased due to two
additional sources of uncertainty that were not considered previously:
the bias correction and the differences of the \deshift distributions
among the fit components.
The systematic correlations between the GLW
parameters $\vec{y}=(\Acpp, \Acpm, \Rcpp, \Rcpm)^T$ are
\CORsyst

\begin{table}[ht]
\center
\caption{Summary of systematic uncertainties.}
\label{tab:syst_summary}
\begin{tabular}{lcccc}
\hline\hline\\[-2.5ex]
Source                             & $A_{\CP+}$	& $A_{\CP-}$	& $R_{\CP+}$	& $R_{\CP-}$	\\ [0.5ex]
\hline
Fixed fit parameters               & $0.004$	& $0.005$	& $0.026$	& $0.022$	\\ 
Peaking background                 & $0.014$	& $0.005$	& $0.017$	& $0.013$	\\ 
Bias correction                    & $0.004$	& $0.004$	& $0.006$	& $0.005$	\\ 
Detector charge asym.              & $0.014$	& $0.014$	& -		& -		\\ 
Opposite-\CP background            & -		& $0.003$	& -		& $0.006$	\\ 
\Rcppm vs. $R_\pm$                 & -		& -		& $0.026$ 	& $0.023$	\\ [0.5ex]
Signal self cross-feed             & $0.000$	& $0.001$	& -		& -		\\ 
$\varepsilon(\pi)/\varepsilon(K)$  & -		& -		& $0.009$	& $0.008$	\\ 
\deshift PDFs                      & $0.007$	& $0.011$	& $0.029$	& $0.024$	\\
\hline
\bf{Total}                         & $0.022$	& $0.020$	& $0.051$	& $0.043$	\\
\hline\hline
\end{tabular}
\end{table}

\section{Results}
\label{sec:results}

The signal yields returned from the fit for each of the $D$ decay mode
under study are listed in Table~\ref{tab:sigyields}. We reconstruct
almost 1000 $B^\pm\to D_{\CP} K^\pm$ decays and about four times more 
$B^\pm\to DK^\pm$, $D\to K\pi$ decays. 

\begin{table}[!h]
\begin{center}
\caption{Measured signal yields calculated from the fit results given in
Table~\ref{tab:fitresult} using $N(\btdk)=N^{\sigp}_{\rm tot} \Rkp$,
$N(\btdp)\equiv N^{\sigp}_{\rm tot}$, and error propagation neglecting
small correlations.}
\label{tab:sigyields}
\begin{tabular}{lcc}
\hline\hline\\[-2.5ex]
\Dz mode	& $N(B^\pm\to DK^\pm)$	& $N(B^\pm\to D\pi^\pm)$    \\
\hline
\kk	  	& $  367 \pm 27   $	& $  4091 \pm 70  $	    \\
\pipi	  	& $  110 \pm  9   $	& $  1230 \pm 41  $	    \\
\kspiz	  	& $  338 \pm 24   $	& $  4182 \pm 73  $	    \\
\ksomega  	& $  116 \pm  9   $	& $  1440 \pm 45  $	    \\
\ksphi	  	& $   52 \pm  4   $	& $   648 \pm 27  $	    \\
\kpi	  	& $ 3361 \pm 82   $	& $ 44631 \pm 232 $	    \\
\hline\hline
\end{tabular}
\end{center}
\end{table}

\begin{figure}[!htb]
\begin{center}
\includegraphics[width=0.35\textwidth]{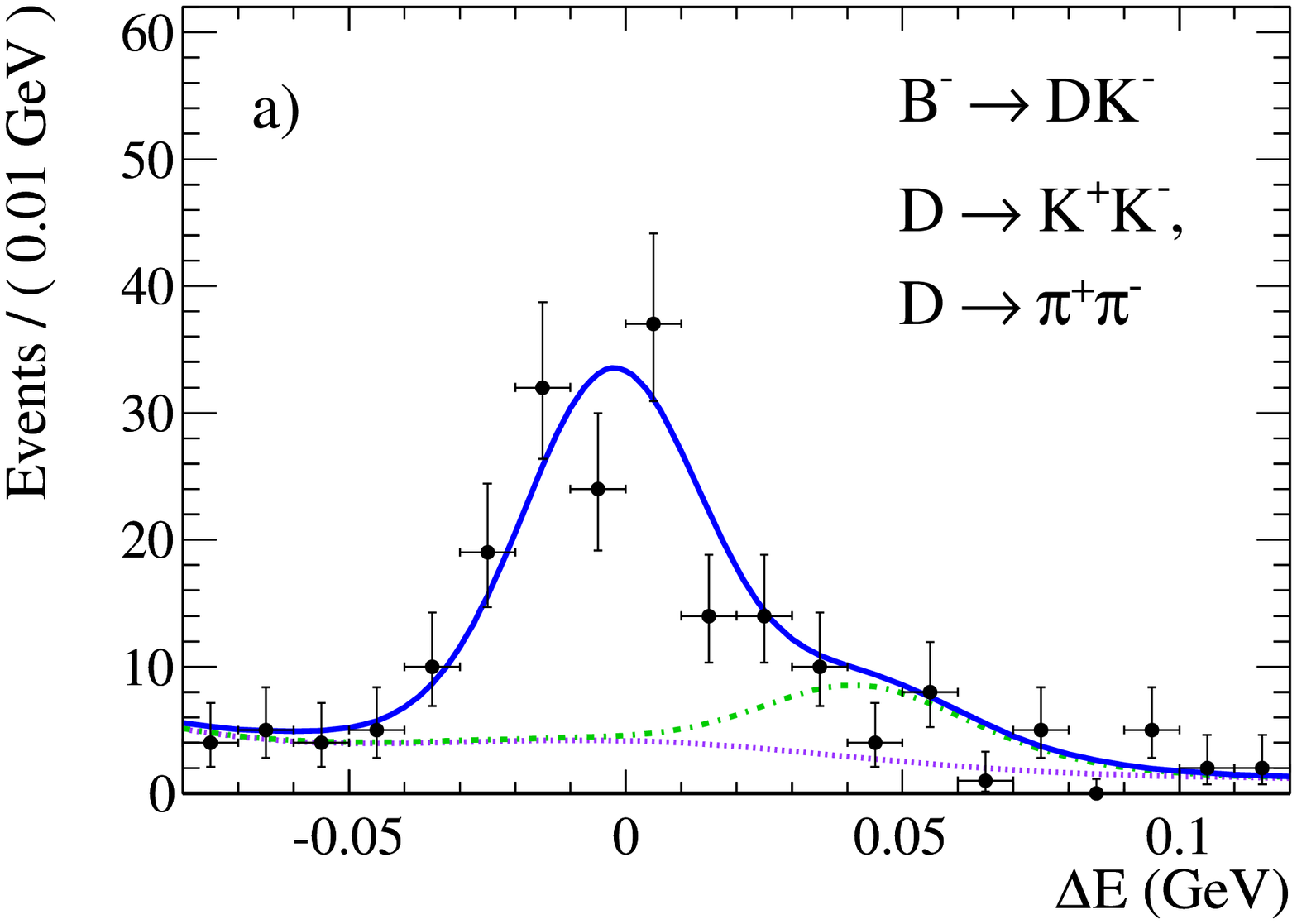}
\includegraphics[width=0.35\textwidth]{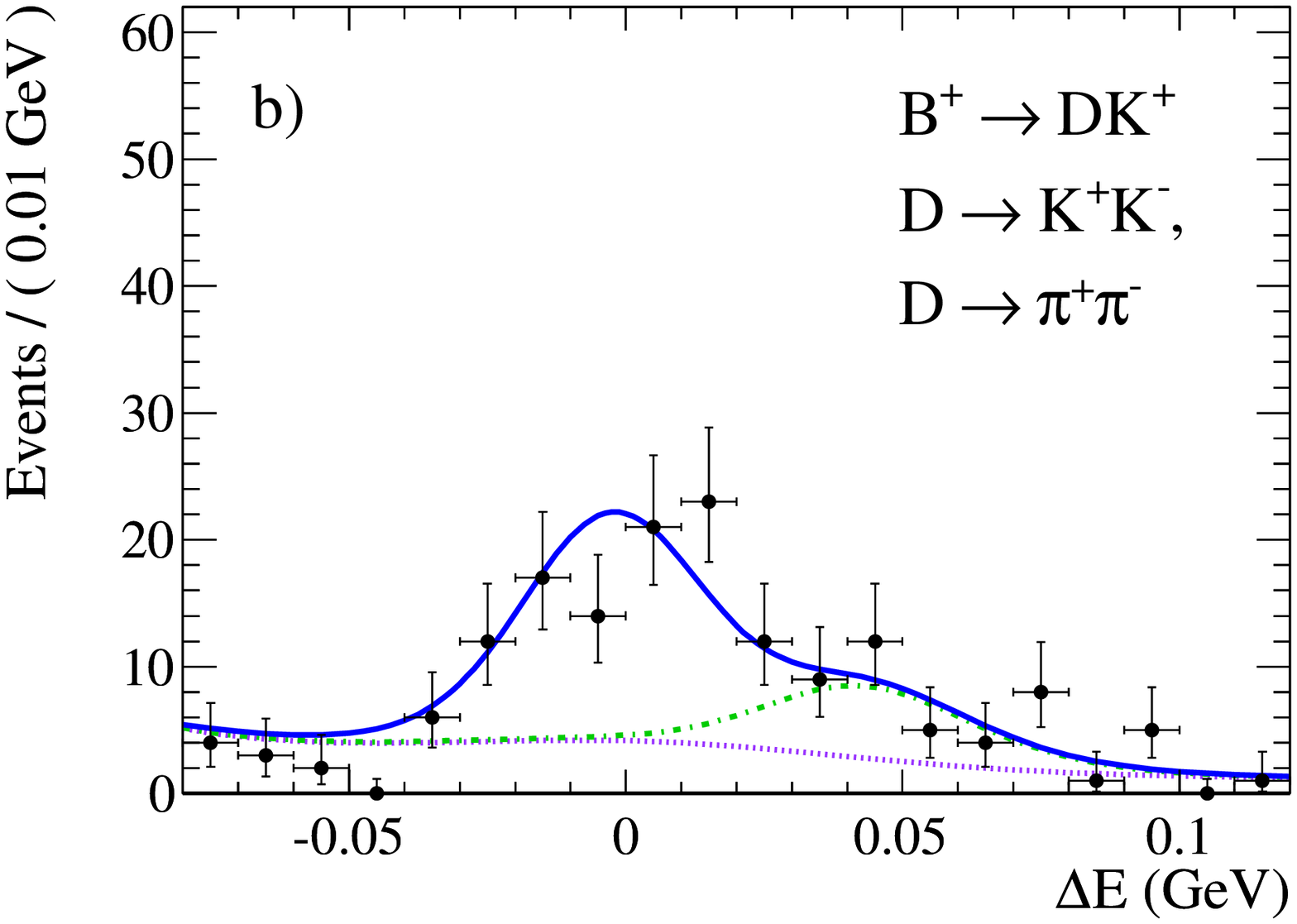}
\includegraphics[width=0.35\textwidth]{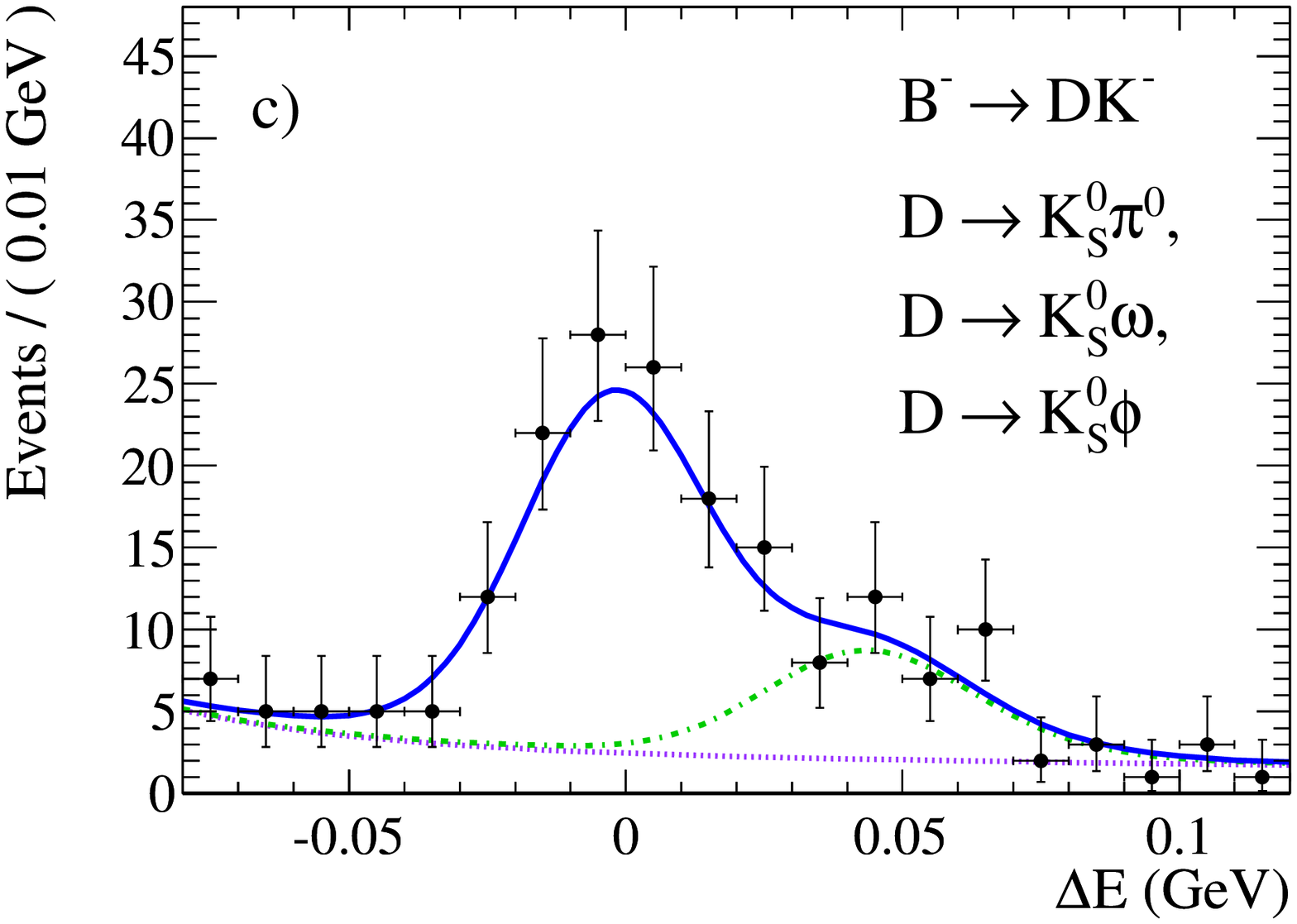}
\includegraphics[width=0.35\textwidth]{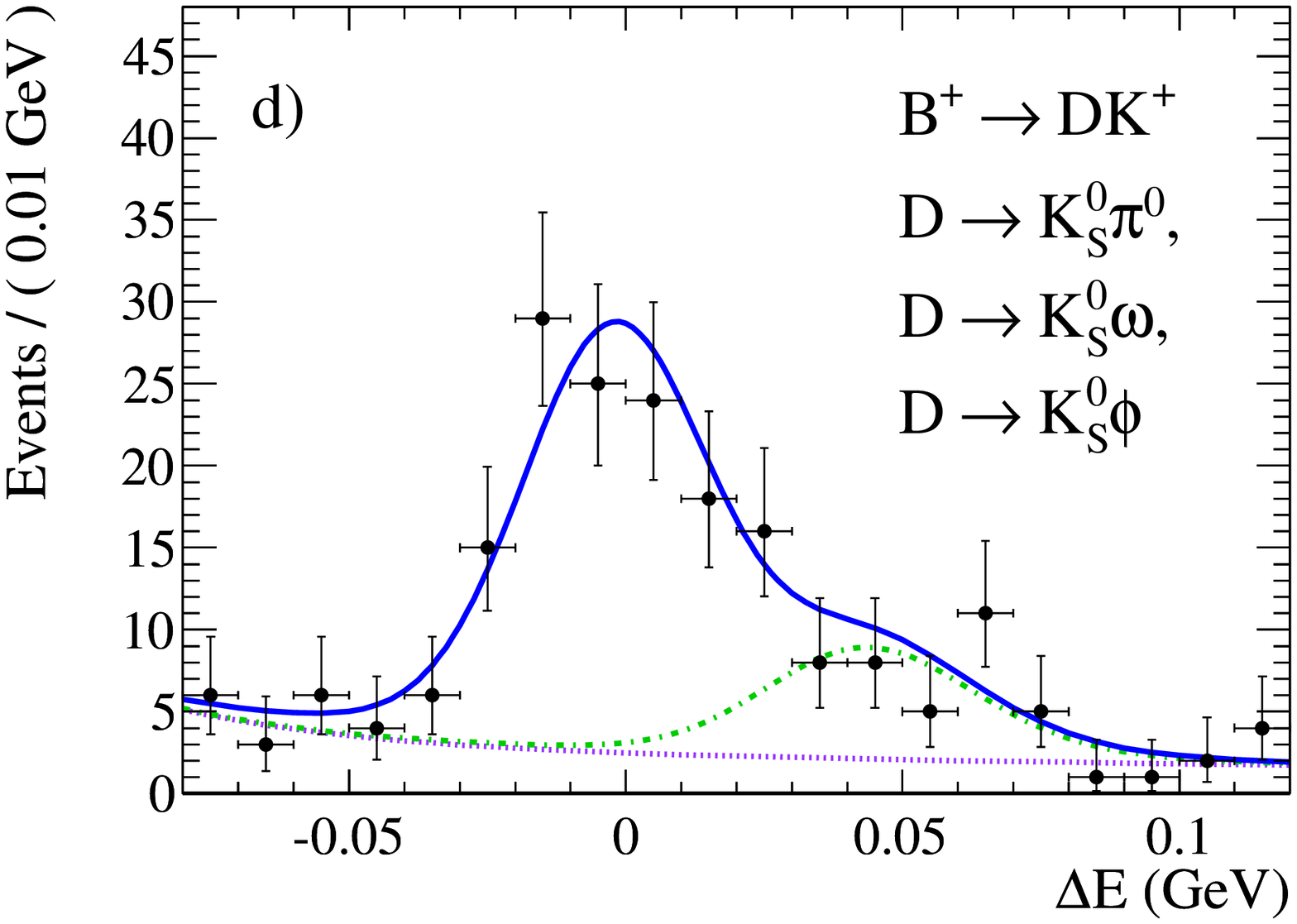}
\caption{\de projections of the fits to the data, split into subsets of definite \CP of the $D$ candidate and charge of the $B$ candidate:
a) $\Bm{\to}D_{\CP+}\Km$, 
b) $\Bp{\to}D_{\CP+}\Kp$, 
c) $\Bm{\to}D_{\CP-}\Km$, 
d) $\Bp{\to}D_{\CP-}\Kp$.
The curves are the full PDF (solid, blue), and $B{\to}D\pi$ (dash-dotted, green) stacked on the remaining backgrounds
(dotted, purple). The region between the solid and the dash-dotted
lines represents the $B{\to}DK$ contribution.
We show the subsets of the data sample in which the track 
$h$ from the $B$ decay is identified as a kaon. 
We require candidates to lie inside the signal-enriched region
defined in Sec.~\ref{sec:selection}, except for the plotted variable.}
\label{fig:data-cpp-cpm}
\end{center}
\end{figure}

\begin{figure}[!htb]
\begin{center}
\includegraphics[width=0.35\textwidth]{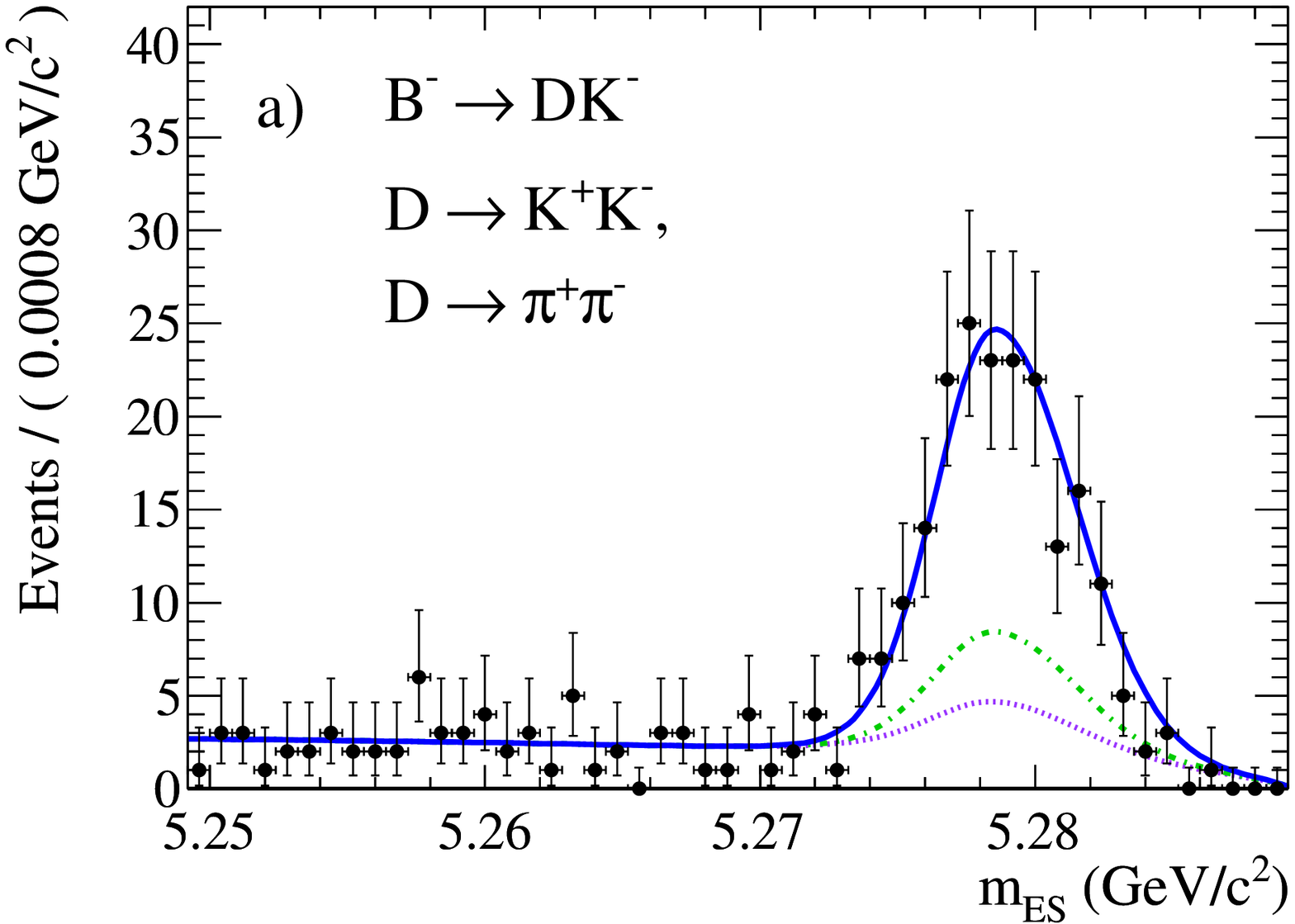}
\includegraphics[width=0.35\textwidth]{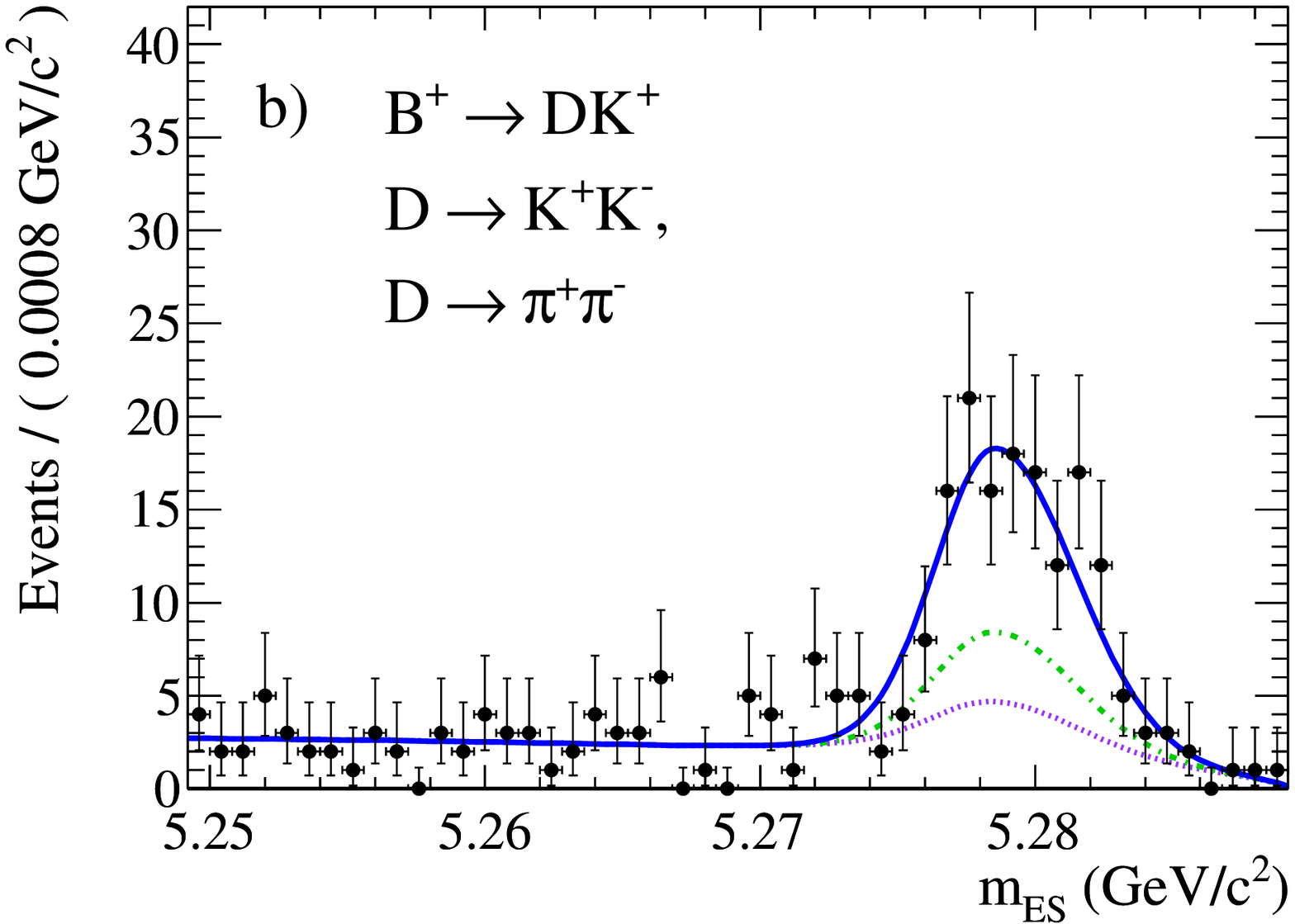}
\includegraphics[width=0.35\textwidth]{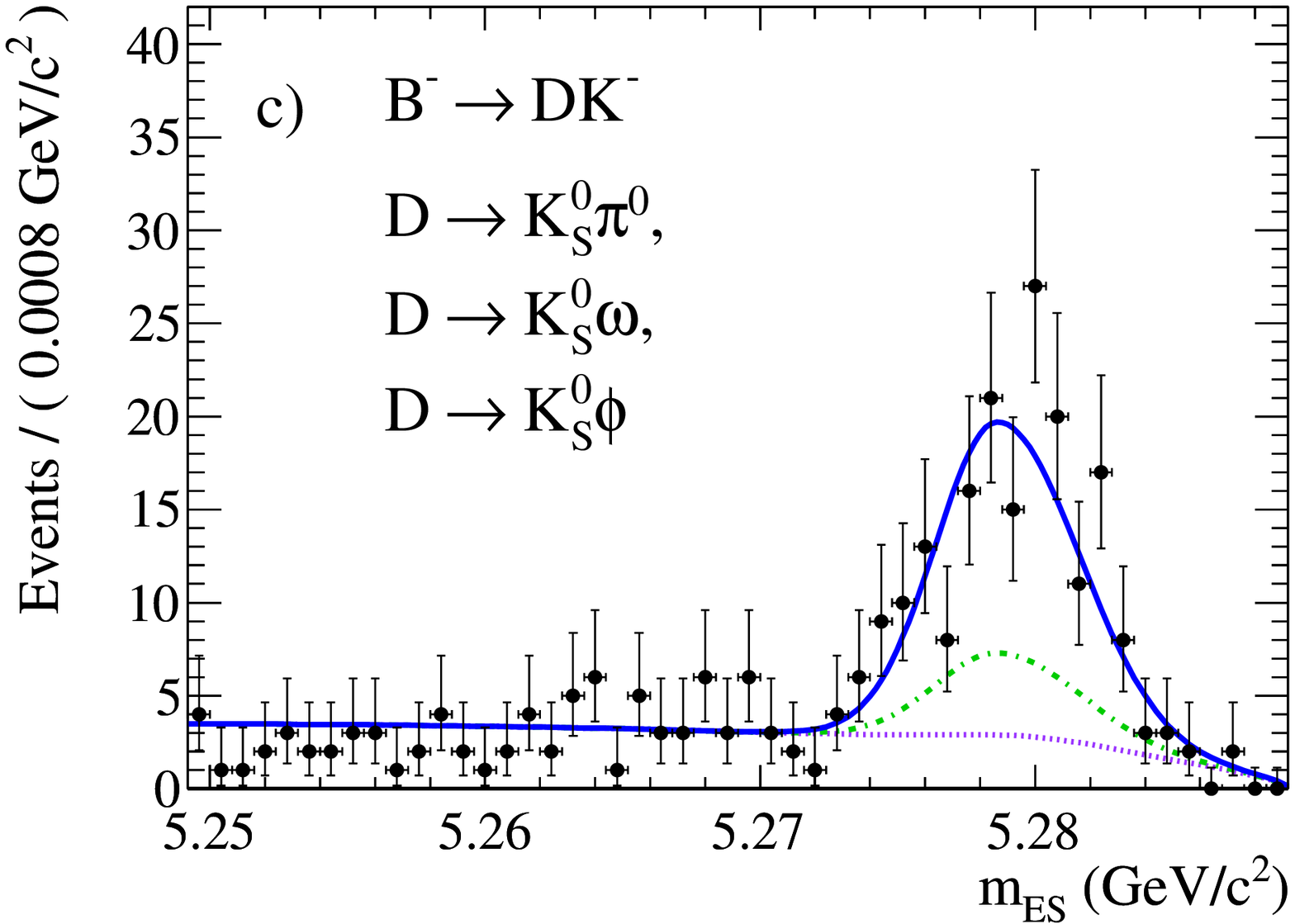}
\includegraphics[width=0.35\textwidth]{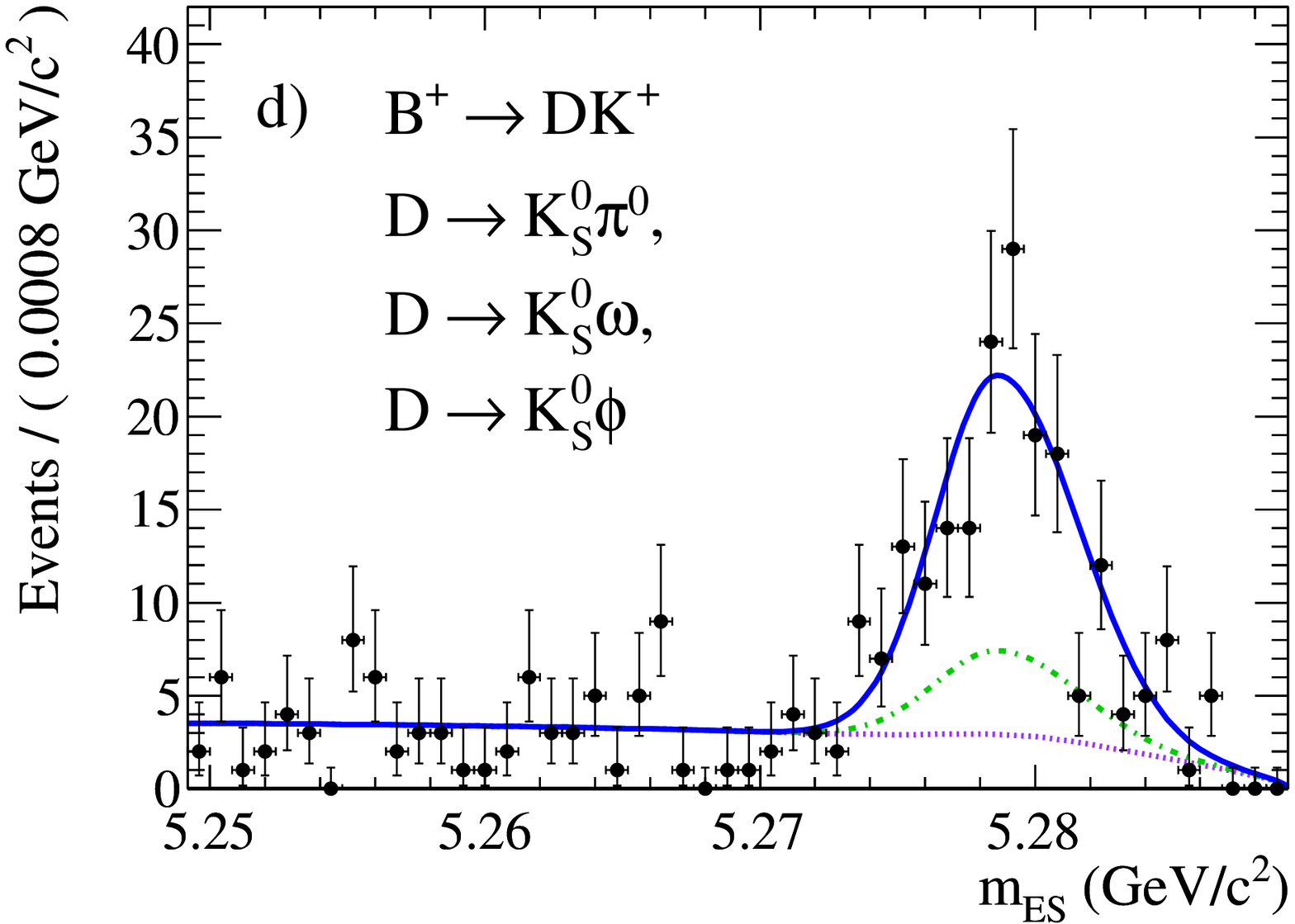}
\caption{\mes projections of the fits to the data, split into subsets of definite \CP of the $D$ candidate and charge of the $B$ candidate:
a) \bmtodcppk, 
b) \bptodcppk, 
c) \bmtodcpmk, 
d) \bptodcpmk.
We show the subsets of the data sample in which the track $h$ from the $B$ decay is identified as a kaon.
See caption of Fig.~\ref{fig:data-cpp-cpm} for line definitions.
Only a subrange of the whole fit range is shown in order to provide a
closer view of the signal peak.
}
\label{fig:data-mes}
\end{center}
\end{figure}

\begin{figure}[!htb]
\begin{center}
\includegraphics[width=0.35\textwidth]{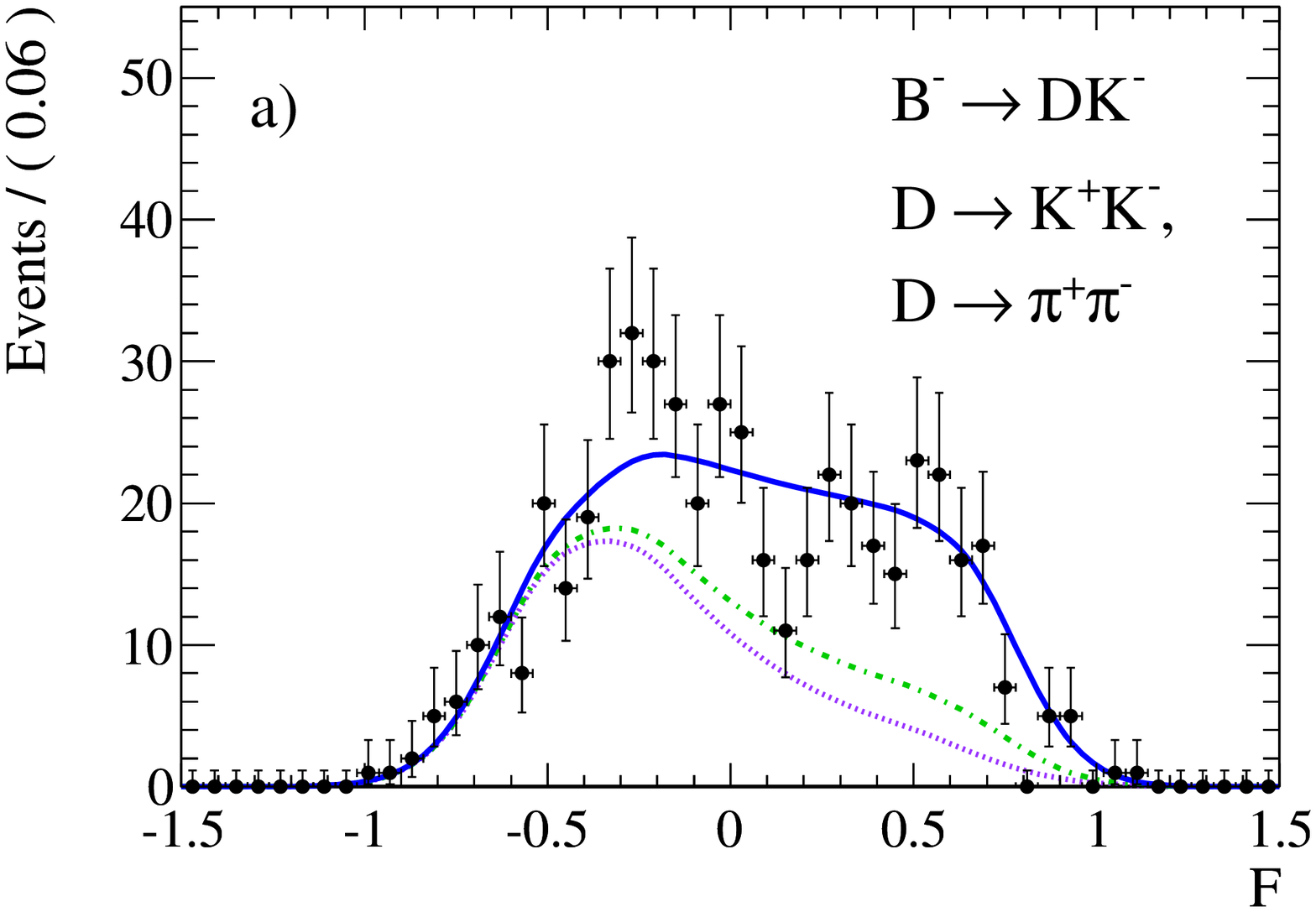}
\includegraphics[width=0.35\textwidth]{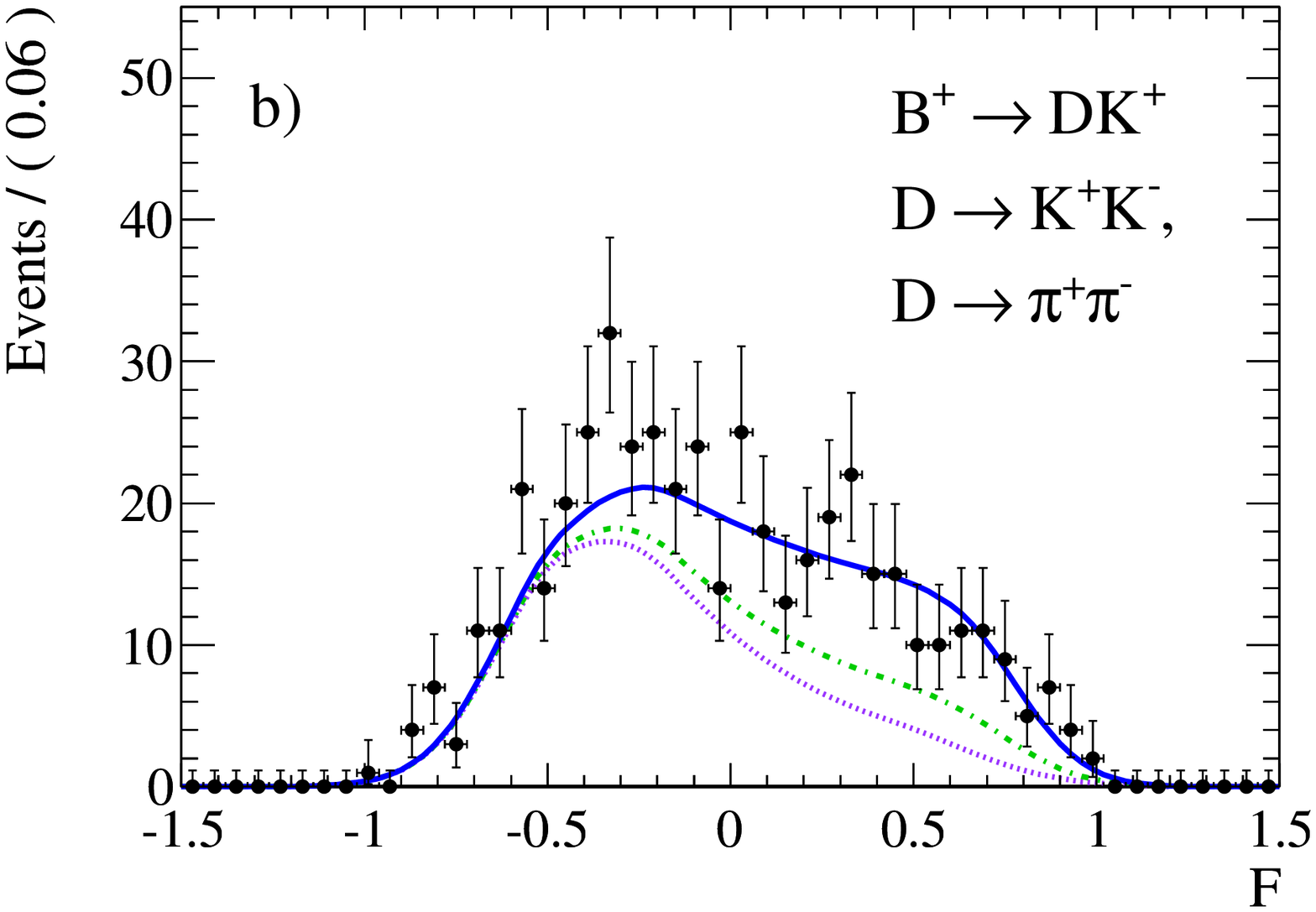}
\includegraphics[width=0.35\textwidth]{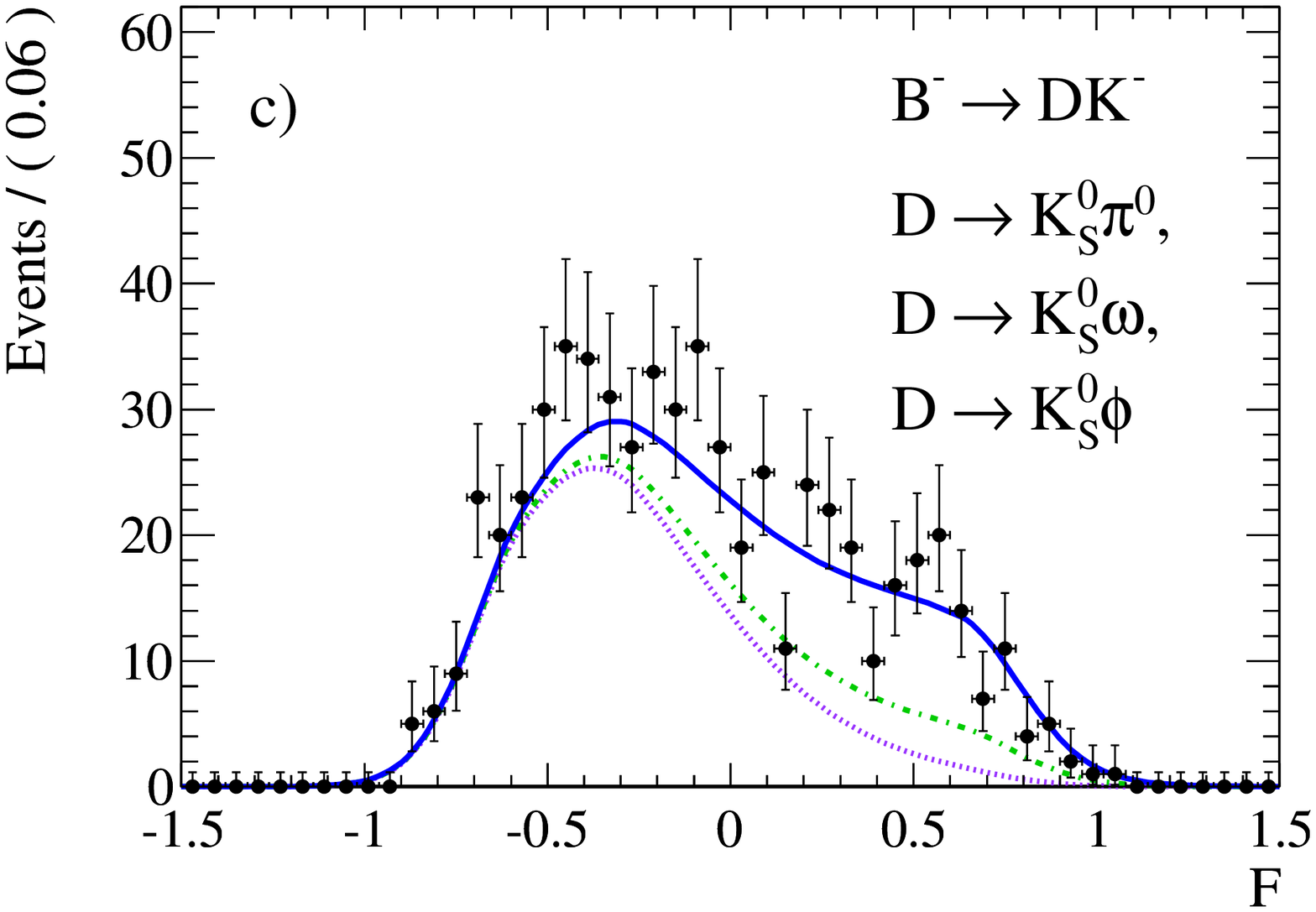}
\includegraphics[width=0.35\textwidth]{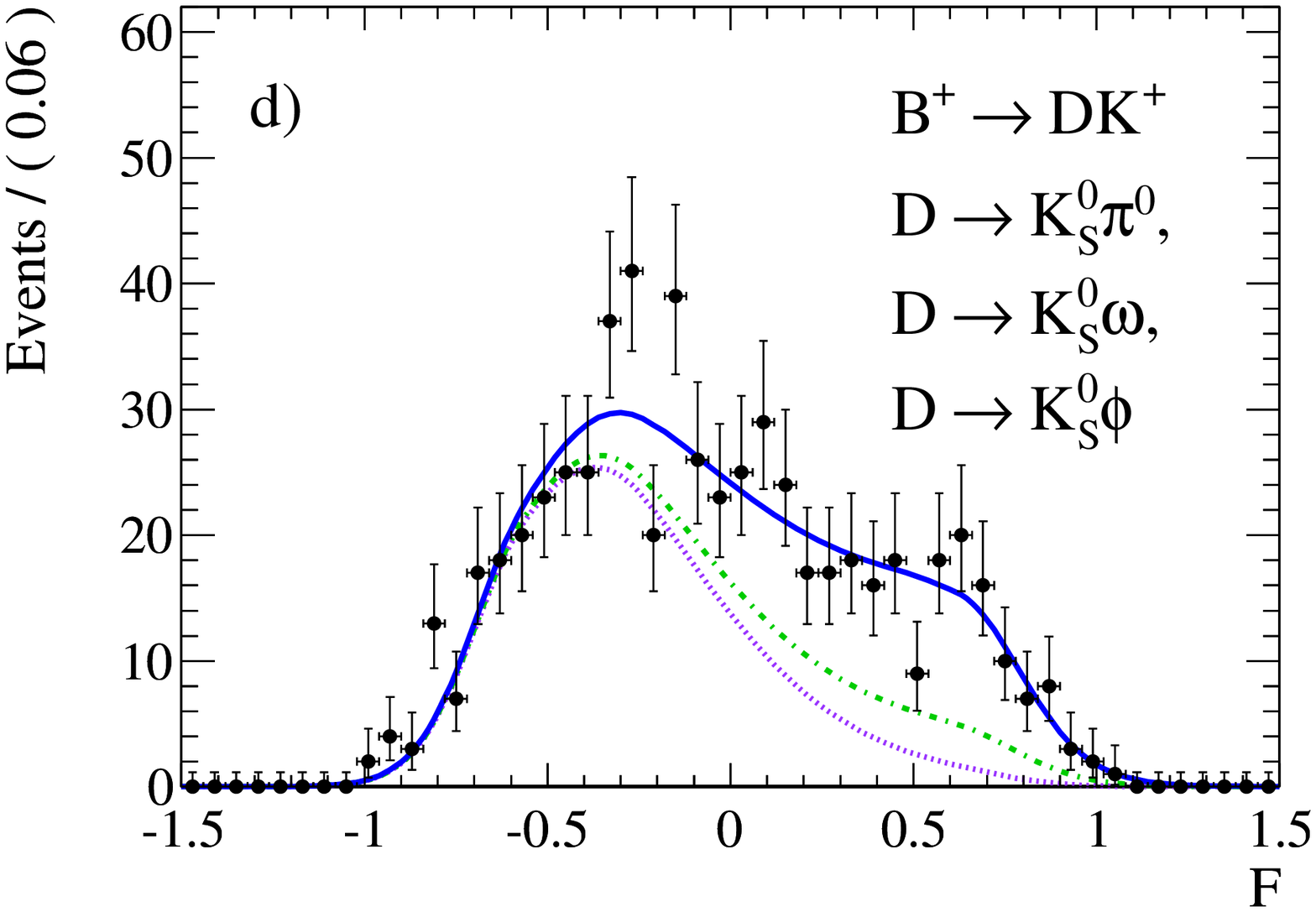}
\caption{\fshr projections of the fits to the data, split into subsets of definite \CP of the $D$ candidate and charge of the $B$ candidate:
a) \bmtodcppk,
b) \bptodcppk, 
c) \bmtodcpmk, 
d) \bptodcpmk.
We show the subsets of the data sample in which the track $h$ from the $B$ decay is identified
as a kaon.
See caption of Fig.~\ref{fig:data-cpp-cpm} for line definitions.}
\label{fig:data-fshr}
\end{center}
\end{figure}

\begin{figure}[!htb]
\begin{center}
\includegraphics[width=0.38\textwidth]{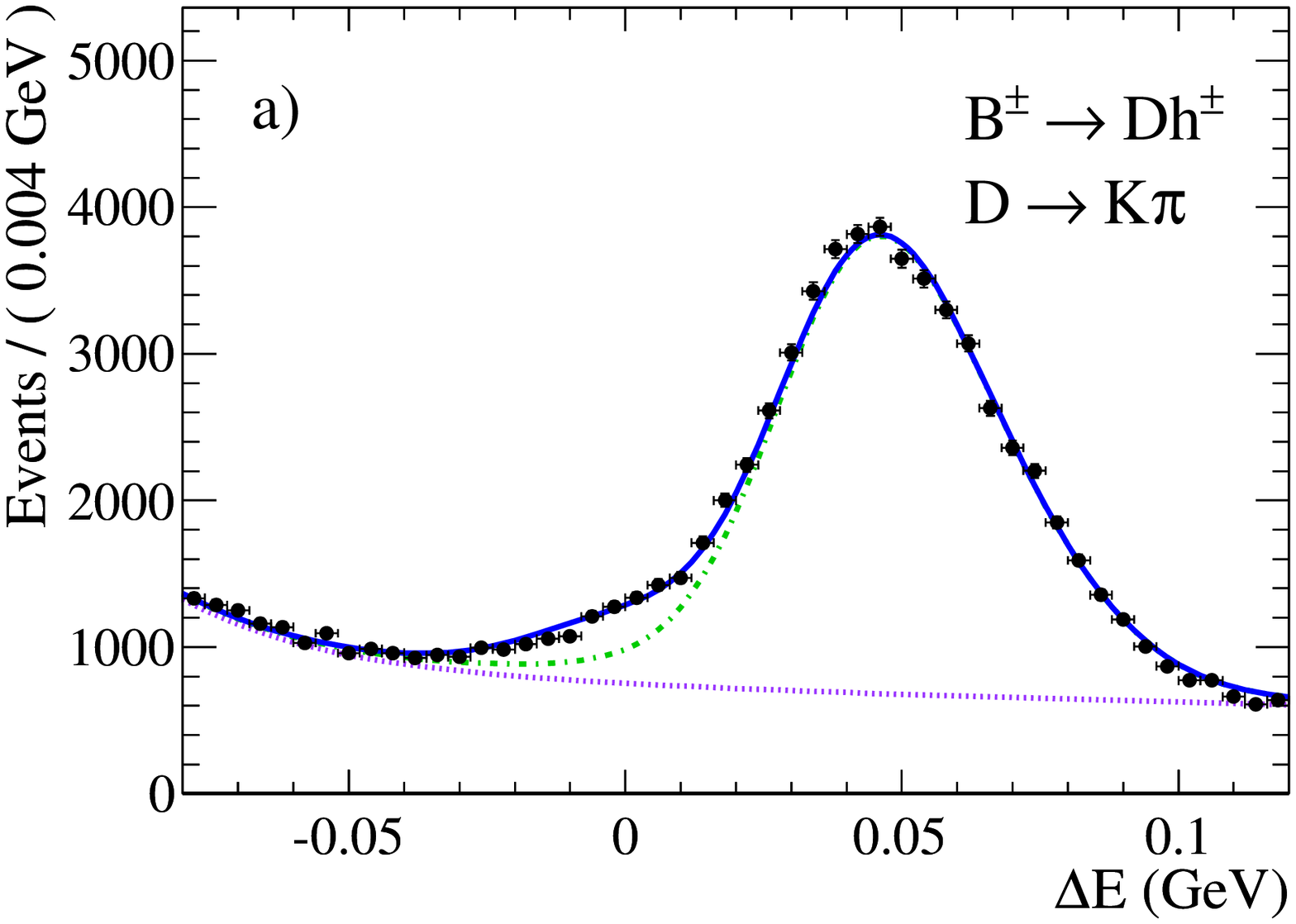}
\includegraphics[width=0.38\textwidth]{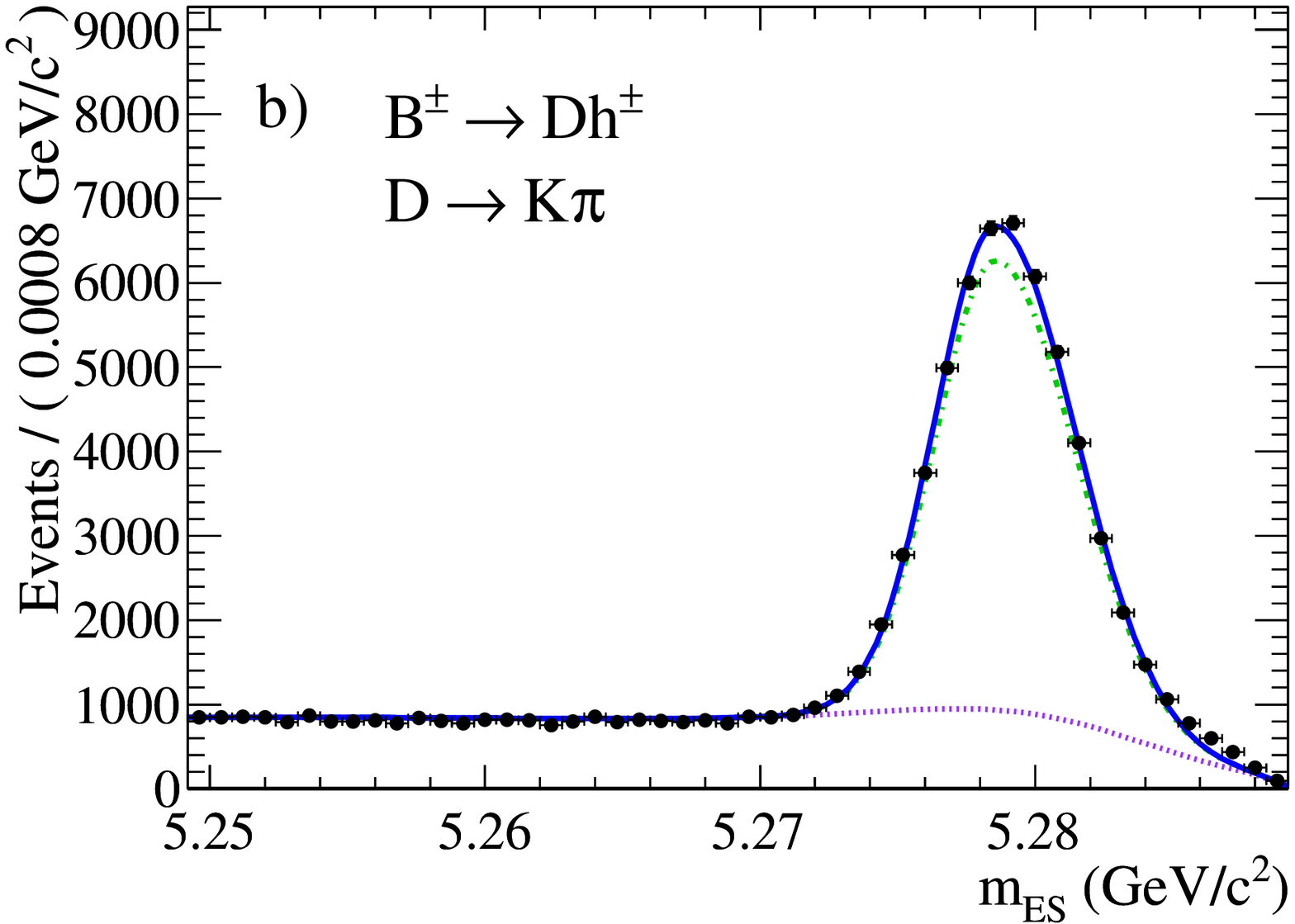}
\includegraphics[width=0.38\textwidth]{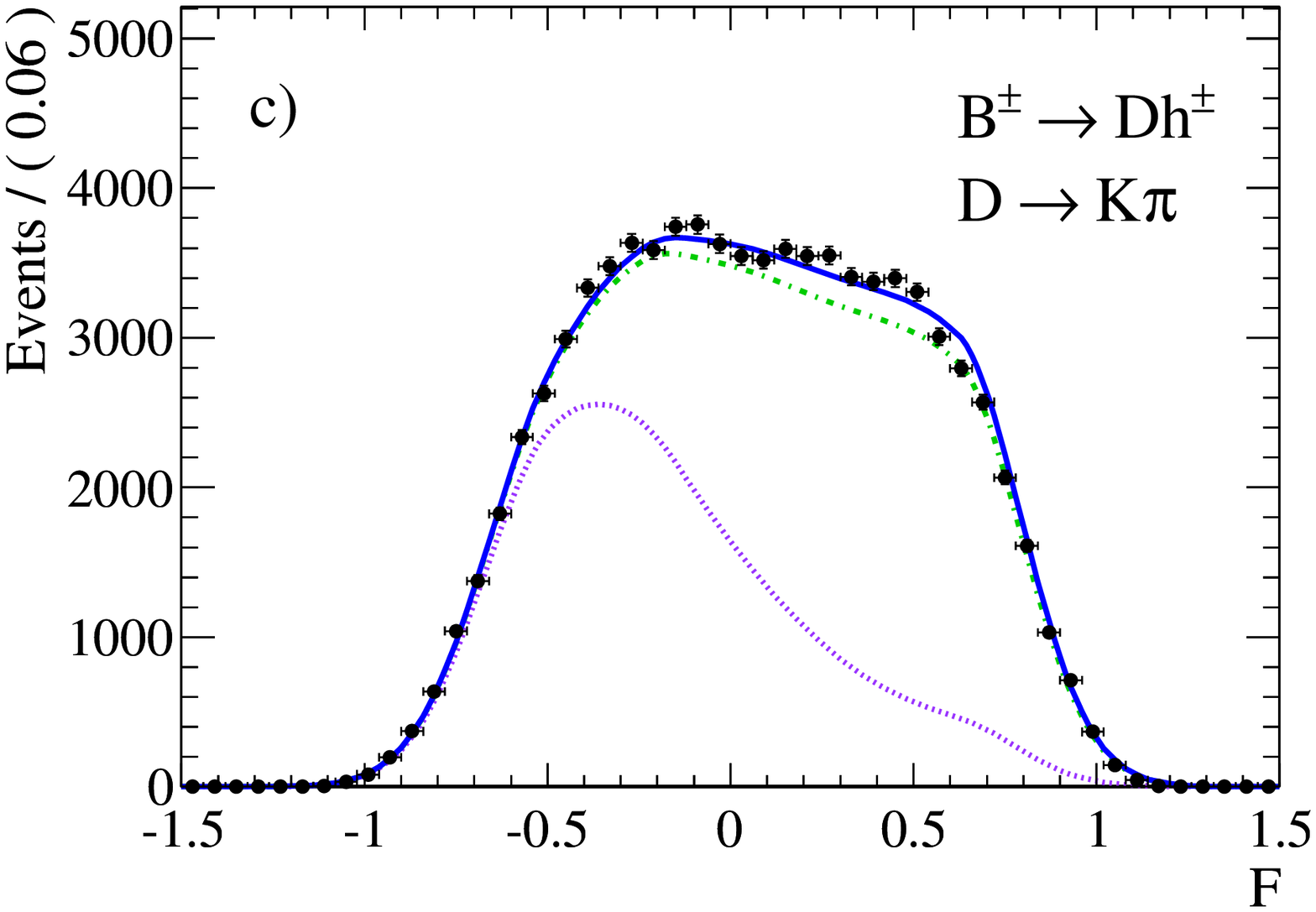}
\caption{Projections of (a) \de, (b) \mes, and (c) \fshr variables
of the fit to the $B^\pm\to Dh^\pm$, \dztokpi flavor mode. No requirements
are put on the PID of the track $h$ from the $B$ decay and on the fit variables not plotted.
See caption of Fig.~\ref{fig:data-cpp-cpm} for line definitions.}
\label{fig:data-kpi}
\end{center}
\end{figure}

The final values of the GLW parameters that we measure are:
\begin{eqnarray}
  \label{eq:result}
  \Acpp &=& \phantom{-}\AcppVal\pm\AcppErrStat\stat\pm\AcppErrSyst\syst\,,\\
  \Acpm &=&            \AcpmVal\pm\AcpmErrStat\stat\pm\AcpmErrSyst\syst\,,\\
  \Rcpp &=& \phantom{-}\RcppVal\pm\RcppErrStat\stat\pm\RcppErrSyst\syst\,,\\
  \Rcpm &=& \phantom{-}\RcpmVal\pm\RcpmErrStat\stat\pm\RcpmErrSyst\syst\,.
\end{eqnarray}
The statistical correlations among these four quantities are:
\CORstat
The results are in good agreement with those from our previous
analysis~\cite{babar_d0k_GLW_PRD} and the current world averages~\cite{HFAG}. 
Figure~\ref{fig:data-cpp-cpm} shows the \de projections of the final
fits to the \CP subsamples and Figures~\ref{fig:data-mes}-\ref{fig:data-kpi}
show \mes and \fshr projections as well as projections of the fit to
the \dztokpi flavor mode.

The statistical significance of a non-zero \Acpp value
is determined from the maximum value of the likelihood function of the
nominal fit and that of a dedicated null-hypothesis fit, where \Acpp was fixed
to zero,
\begin{equation}
	S_{\rm stat} = \sqrt{2\ln(\mathcal{L}_{\rm nom}/\mathcal{L}_{\rm null})} = \sigStat.
\end{equation}
Taking into account systematic uncertainties, 
the statistical significance of $A_{\CP+}$ is slightly decreased to:
\begin{equation}
	S_{{\rm stat} + {\rm syst}} = \frac{S_{\rm stat}}{\sqrt{1+\frac{\sigma_{\rm syst}^2}{\sigma_{\rm stat}^2}}} = \sigStatSyst.
\end{equation}
This constitutes evidence for direct \CP violation in charged $B$
decays and the first evidence of direct \CP violation in $B\to DK$.

\begin{figure}[!htb]
\center
\includegraphics[width=0.38\textwidth]{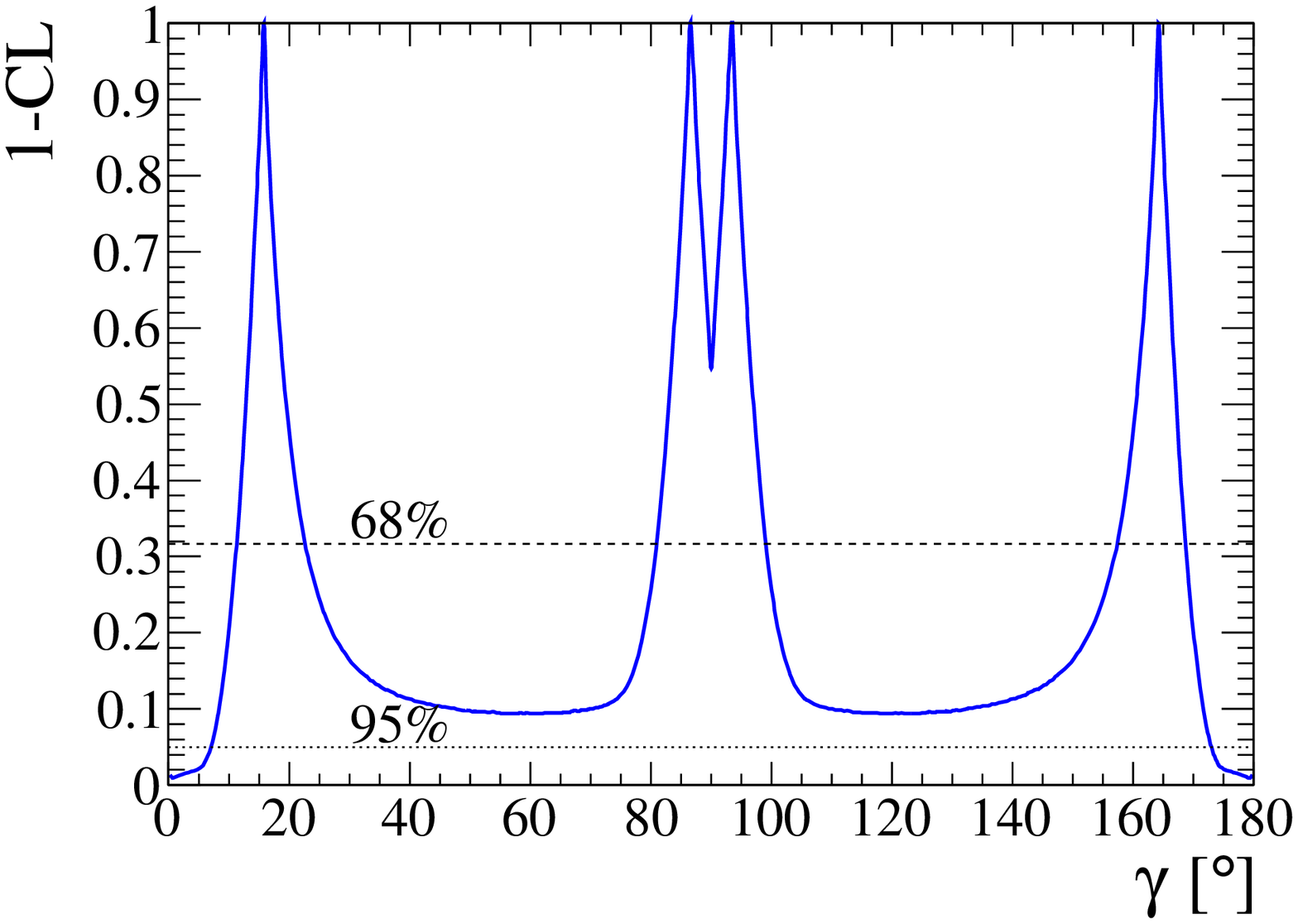}
\includegraphics[width=0.38\textwidth]{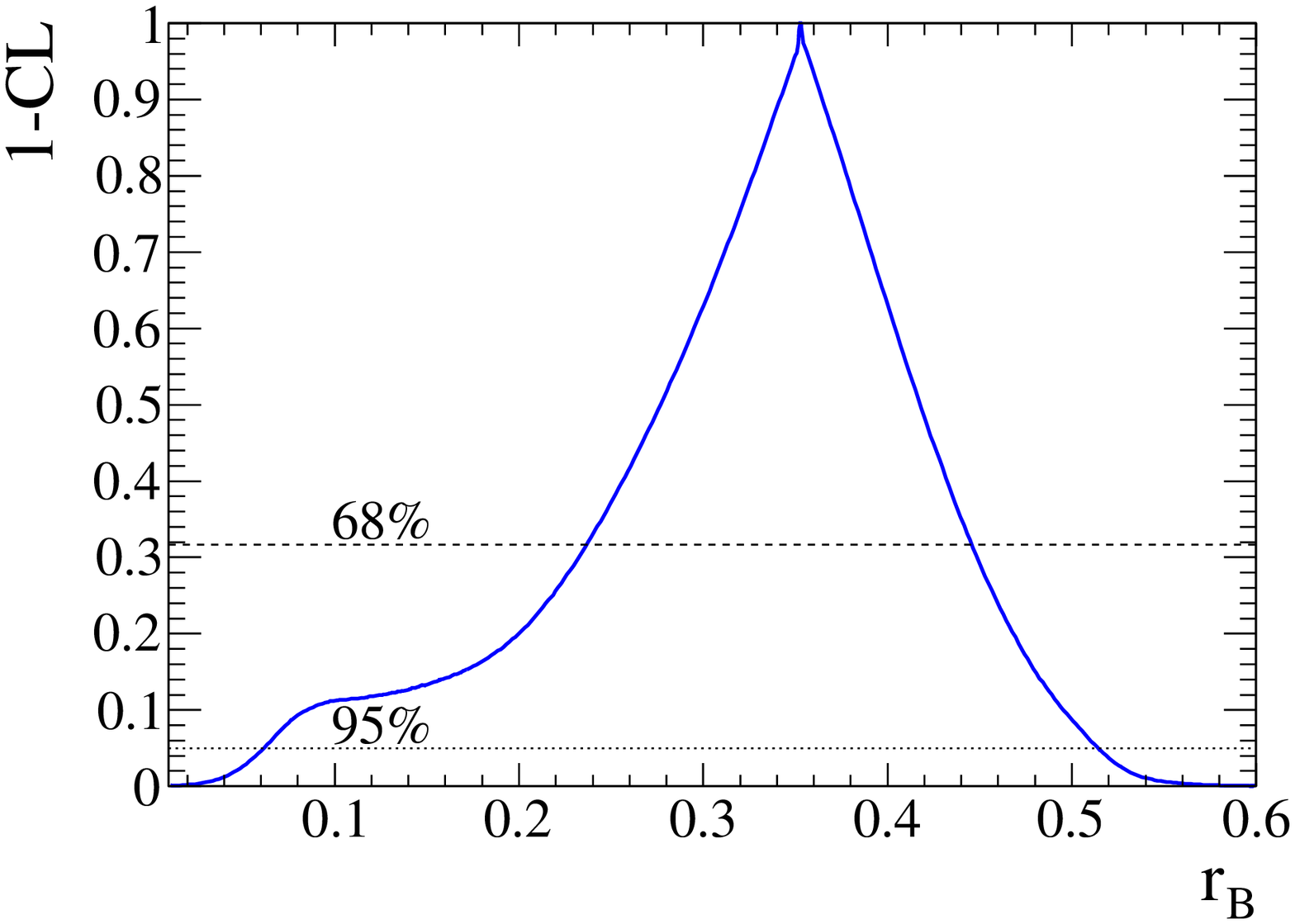}
\caption{1-CL as a function of $\gamma$ (top) and $r_B$ (bottom). Both statistical
and systematic uncertainties are taken into account. For the angle
$\gamma$, the plot is identical in the range $[180\degrees,360\degrees]$.
The horizontal
lines show the 68\% CL (dashed) and the 95\% CL (dotted). Due to the
symmetry of Eqns.~\ref{eq:rcp} and~\ref{eq:acp} the plot for the strong
phase $\delta_B$ is identical to the one for $\gamma$.}
\label{fig:cl1d}
\end{figure}

\begin{figure}[!htb]
\center
\includegraphics[width=0.38\textwidth]{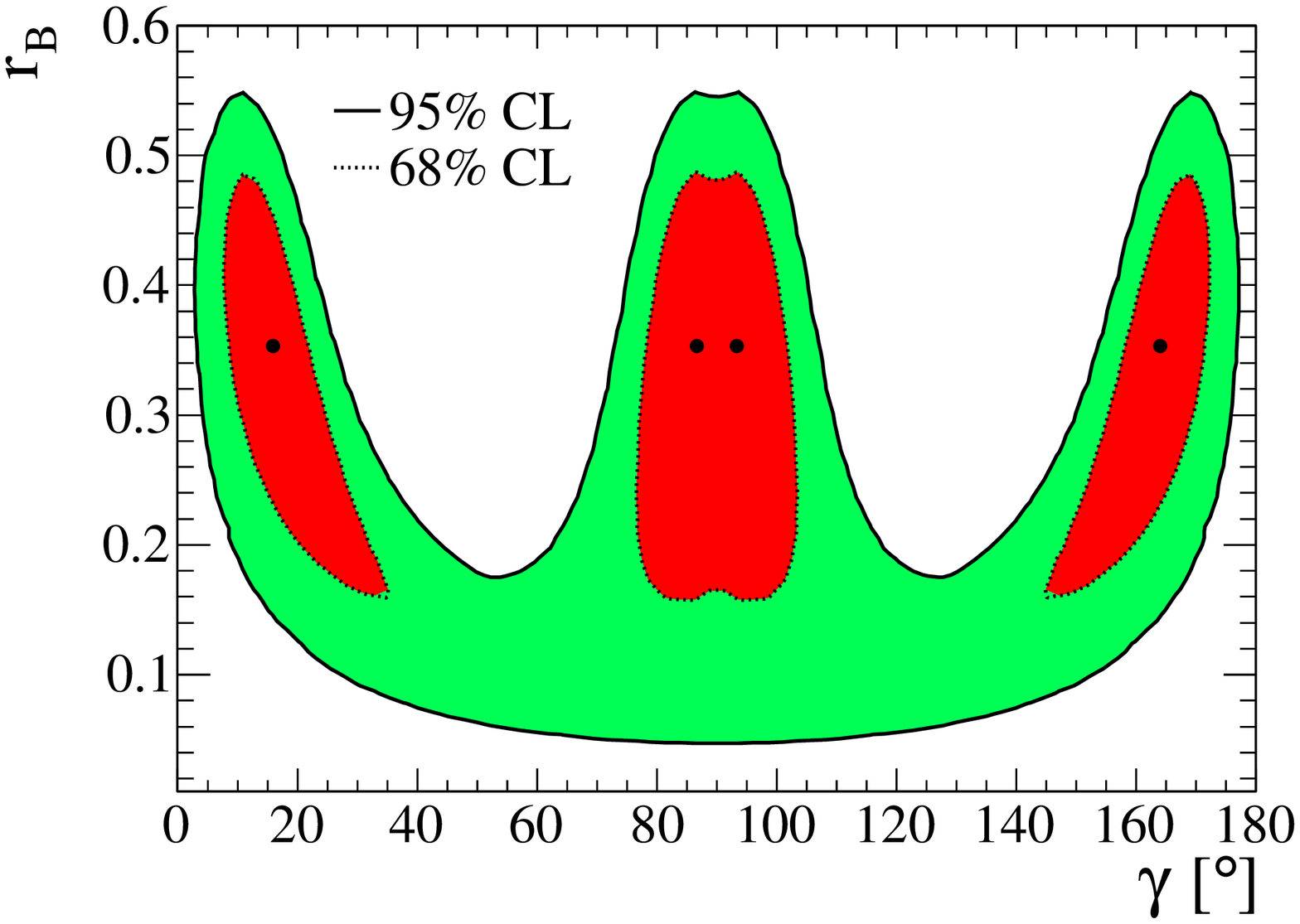}
\includegraphics[width=0.38\textwidth]{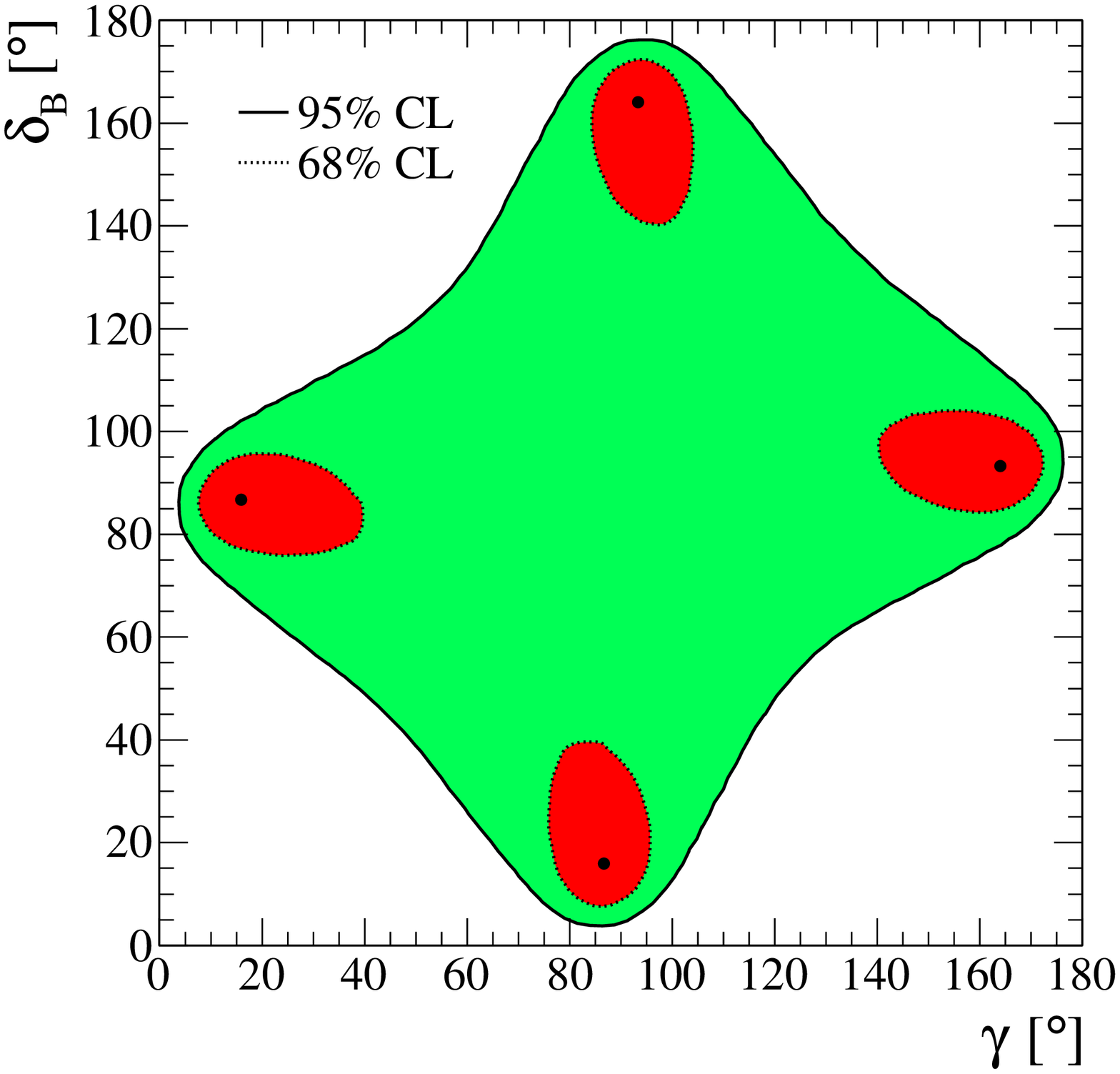}
\caption{Contours at 68\% (dotted, red) and 95\% (solid,
green) 2-dimensional CL in the $(\gamma, r_B)$ and $(\gamma, \delta_B)$ planes.
See also the caption of Fig.~\ref{fig:cl1d} regarding symmetries.}
\label{fig:cl2d}
\end{figure}

We constrain the CKM angle $\gamma$, the strong phase $\delta_B$, and the
amplitude ratio $r_B$ from the present measurement by
adopting the frequentist procedure also exploited in~\cite{babar_dk_DALITZ_PRD}.
We define a multivariate Gaussian likelihood function
\begin{equation}
	\label{eq:gammaLikelihood}
	\mathcal{L}(\gamma,\delta_B,r_B) = \frac{1}{N} \exp\left(-\frac{1}{2} (\vec{y}-\vec{y_{\rm t}})^T V_{\rm cov}^{-1} (\vec{y}-\vec{y_{\rm t}})\right)
\end{equation}
relating the experimentally measured observables $\vec{y}$ and their
statistical and systematic covariance matrices $V_{\rm cov} = V_{\rm stat}+V_{\rm
syst}$ with the corresponding truth parameters $\vec{y_{\rm t}}=\vec{y_{\rm
t}}(\gamma,\delta_B,r_B)$ calculated using Eqns.~\ref{eq:rcp}
and~\ref{eq:acp}. The matrices $V_{\rm stat}$ and $V_{\rm syst}$ are
constructed from Eqns.~\ref{eq:cor_syst}-\ref{eq:cor_stat}. The
normalization is $N=(2\pi)^2\sqrt{|V_{\rm cov}|}$. We then define a 
$\chi^2$-function as
\begin{equation}
	\label{eq:chi2result}
	\chi^2(\gamma,\delta_B,r_B) = -2 \ln \mathcal{L}(\gamma,\delta_B,r_B).
\end{equation}
Due to the inherent eight-fold ambiguity of the GLW method there are
eight equivalent minima of the $\chi^2$-function, $\chi^2_{\min}$,
which correspond to the same value of $r_B$ and to eight alternative
solutions for $(\gamma,\delta_B)$. To
evaluate the confidence level of a certain truth parameter (for
example $\gamma$) at a certain value ($\gamma_0$) we consider the value
of the $\chi^2$-function at the new minimum, $\chi^2_{\min}(\gamma_0,
\delta_B', r_B')$, satisfying $\Delta\chi^2 = \chi^2_{\min}(\gamma_0,
\delta_B', r_B') - \chi^2_{\min} \ge 0$. In a purely Gaussian situation
for the truth parameters the CL is given by the probability that
$\Delta\chi^2$ is exceeded for a $\chi^2$-distribution with one degree
of freedom:
\begin{equation}
	\label{eq:cl}
	{1-\rm CL} = \frac{1}{\sqrt{2} \, \Gamma(1/2)} \int_{\Delta\chi^2}^\infty e^{-t/2} t^{-1/2} \, {\rm d}t.
\end{equation}
A more accurate approach is to take into account the non-linearity of
the GLW relations, Eqns.~\ref{eq:rcp} and~\ref{eq:acp}.
In this case one should consider $\Delta\chi^2$ as a
test statistic, and calculate $(1-\rm CL)$ by means of a Monte Carlo
procedure, described in the following. For a certain value of interest
$(\gamma_0)$, we:
\begin{enumerate}
  \item calculate $\Delta\chi^2 = \chi^2_{\min}(\gamma_0, \delta_B', r_B')
  - \chi^2_{\min}$ as before;
  \item generate a ``toy'' result $\Acppm'$, $\Rcppm'$, using
  Eq.~\ref{eq:gammaLikelihood} with values $\gamma_0$, $\delta_B'$, $r_B'$
  as the PDF;
  \item calculate $\Delta\chi^{2\prime}$ of the toy result as in
  the first step, \emph{i.e.} minimize again with respect to $\delta_B$ and $r_B$;
  \item calculate $(1-\rm CL)$ as the fraction of toy results which
  perform better than the measured data, \textit{i.e.} $1-\rm CL = N(\Delta\chi^2 >
  \Delta\chi^{2\prime})/N_{\rm toy}$.
\end{enumerate}
Figures~\ref{fig:cl1d} and~\ref{fig:cl2d} illustrate 1-CL as a
function of $\gamma$ and $r_B$ as obtained from this study. From these
distributions we extract 68\% and 95\% CL confidence intervals for
$\gamma$ and $r_B$, as summarized in Table~\ref{tab:cl_toy}.
Due to the $\gamma\leftrightarrow\delta_B$ ambiguity of the GLW method, the 1D CL 
intervals for $\delta_B$ are identical to those for $\gamma$.
At the 68\% CL we are able to distinguish six out of
eight solutions for $\gamma$ (and $\delta_B$), two of which are
in good agreement with the current world averages~\cite{PDG2008}. At the
95\% CL we are able to exclude the intervals $[0\degrees, \dgLoBi\degrees]$,
$[\dgHiBi\degrees, \dgLoBj\degrees]$ and $[\dgHiBj\degrees, 360\degrees]$
 for $\gamma$ and $\delta_B$.
For $r_B$ we deduce at 68\% CL:
\begin{equation}
   \label{eq:rb} 
   r_B = \rbVal^{\rbErrHi}_{\rbErrLo}\sysstat.
\end{equation}

\begin{table}[!htb]
\center
\caption{68\% and 95\% CL intervals for the parameters $\gamma$, and
$r_B$, taking into account both statistical and systematic uncertainties. 
The confidence intervals for $\delta_B$ are identical to those for
$\gamma$ due to the intrinsic
$\gamma\leftrightarrow\delta_B$ ambiguity of the GLW method.}
\label{tab:cl_toy}
\begin{tabular}{lcc}
\hline\hline
                 & $\gamma\mod 180\,[{}^{\circ}]$ & $r_B$ \\
\hline
68\% CL    	     & $[\dgLoAi, \dgHiAi]$ & $[\rLoA, \rHiA]$ \\
                 & $[\dgLoAj, \dgHiAj]$ & \\
		         & $[\dgLoAk, \dgHiAk]$ & \\
95\% CL   	 & $[\dgLoBi,  \dgHiBi]$  & $[\rLoB, \rHiB]$ \\
\hline\hline
\end{tabular}
\end{table}

In order to facilitate the future combination of these measurements
with the results of the Dalitz plot analysis of $B^\pm\to DK^\pm$, $D \to \KS
h^+ h^-$ decays ($h=\pi,K$)~\cite{babar_dk_DALITZ_PRL}, 
we recompute the GLW parameters after excluding
from the nominal fit the $D_{\CP-}\to\ksphi$ ($\phi\to\KpKm$)
subsample. The sample obtained in this way is statistically
independent of that selected in~\cite{babar_dk_DALITZ_PRL}.
The final values of the GLW parameters that we measure in this case are:
\begin{eqnarray}
  \label{eq:result_noksphi}
  \Acpp &=& \phantom{-}\AcppVal\pm\AcppErrStat\stat\pm\AcppErrSyst\syst\,,\\
  \Acpm &=&            \AcpmValNoKsPhi\pm\AcpmErrStatNoKsPhi\stat\pm\AcpmErrSystNoKsPhi\syst\,,\\
  \Rcpp &=& \phantom{-}\RcppVal\pm\RcppErrStat\stat\pm\RcppErrSyst\syst\,,\\
  \Rcpm &=& \phantom{-}\RcpmValNoKsPhi\pm\RcpmErrStatNoKsPhi\stat\pm\RcpmErrSystNoKsPhi\syst\,.
\end{eqnarray}
The statistical correlations among these four quantities are:
\CORstatNoKsphi
and the systematic correlations are 
\CORsystNoKsphi
To compare the results obtained after removing the $D_{\CP-}\to\ksphi$ subsample
with those from the $B^\pm\to DK^\pm, D\to\KS h^+h^-$
analyses, which are expressed in terms of the variables $x_\pm =
r_B\cos(\delta_B\pm \gamma)$ and $y_\pm =r_B\sin(\delta_B\pm \gamma)$, 
we use the GLW parameters measured in this way to determine
the quantities $x_\pm$ through the relations:
\begin{equation}
  \label{eq:x}
  x_\pm = \frac{1}{4}\left[R_{\CPp}(1\mp A_{\CPp})-R_{\CPm}(1\mp A_{\CPm})\right]\,.
\end{equation}
We obtain
\begin{eqnarray}
  x_+   &=&            \xpValNoKsPhi\pm\xpErrStatNoKsPhi\stat\pm\xpErrSystNoKsPhi\syst\,,	\label{eq:xp}\\
  x_-   &=& \phantom{-}\xmValNoKsPhi\pm\xmErrStatNoKsPhi\stat\pm\xmErrSystNoKsPhi\syst\,.	\label{eq:xm}
\end{eqnarray}
These results are in good agreement with the current world
averages~\cite{HFAG} and have precision close to the single most
precise measurements~\cite{babar_dk_DALITZ_PRL}.
We also measure $r_B^2$, which provides a constraint on $x_\pm$ and
$y_\pm$ via $r_B^2 = x_\pm^2 + y_\pm^2$, from
\begin{equation}
  \label{eq:r2}
  r_B^2 = \frac{1}{2}(\Rcpp + \Rcpm - 2)\,.
\end{equation}
We determine:
\begin{equation}
  r_B^2 = \rbSqValNoKsPhi\pm\rbSqErrStatNoKsPhi\stat\pm\rbSqErrSystNoKsPhi\syst\,.
\end{equation}
The constraints that could be placed on the quantities $y_\pm$ from these
measurements, by exploiting the relation $r_B^2 = x_\pm^2\pm
y_\pm^2$, are much weaker than those provided by the $B^\pm\to DK^\pm$, $D \to
\KS h^+ h^-$ analysis.

%
As a final check of consistency we consider the quantity $a$,
\begin{equation}
	a = \Acpp\Rcpp + \Acpm\Rcpm\,.
\end{equation}
From Eqns.~\ref{eq:rcp} and~\ref{eq:acp} one expects $a$ to satisfy
$a=0$. We measure $a = \ARARval\pm\ARARerr\sysstat$, which is
compatible with 0.

\section{Summary}
\label{sec:summary}

Using the entire dataset collected by \babar\ at the $e^+e^-$ 
center-of-mass energy close to the \FourS mass,
we have reconstructed $B^\pm\to DK^\pm$ decays, with $D$ mesons
decaying to non-\CP ($K\pi$), \CP-even (\kk, \pipi) and \CP-odd
(\kspiz, \ksphi, \ksomega) eigenstates.

Through an improved analysis method compared to the
previous \babar\ measurement~\cite{babar_d0k_GLW_PRD} and through an
enlarged dataset, corresponding to an increase in integrated luminosity 
at the \FourS peak from 348\invfb to 426\invfb, we
obtain the most precise measurements of the GLW parameters \Acppm and
\Rcppm to date:
\begin{eqnarray*}
	\Acpp &=& \phantom{-}\AcppVal\pm\AcppErrStat\stat\pm\AcppErrSyst\syst\,,\\
	\Acpm &=& \AcpmVal\pm\AcpmErrStat\stat\pm\AcpmErrSyst\syst\,,\\
	\Rcpp &=& \phantom{-}\RcppVal\pm\RcppErrStat\stat\pm\RcppErrSyst\syst\,,\\
	\Rcpm &=& \phantom{-}\RcpmVal\pm\RcpmErrStat\stat\pm\RcpmErrSyst\syst\,.
\end{eqnarray*}

We measure a value of \Acpp which is \sigStatSyst\
standard deviations from zero,
which constitutes the first evidence for direct \CP violation 
in $B\to DK$ decays.

From the measured values of the GLW parameters, we extract confidence
intervals for the CKM angle $\gamma$, the strong phase 
$\delta_B$, and the amplitude ratio $r_B$, using a frequentist
approach, taking into account
both statistical and systematic uncertainties.
At the 68\% CL we find that both $\gamma$ and $\delta_B$
(modulo $180\degrees$) belong to one of the three intervals
$[\dgLoAi\degrees, \dgHiAi\degrees]$, 
$[\dgLoAj\degrees, \dgHiAj\degrees]$ or 
$[\dgLoAk\degrees, \dgHiAk\degrees]$,
and that
\begin{equation*}
r_B \in [\rLoA, \rHiA].
\end{equation*}
At 95\% CL, we exclude the intervals $[0\degrees, \dgLoBi\degrees]$,
$[\dgHiBi\degrees, \dgLoBj\degrees]$ and $[\dgHiBj\degrees, 360\degrees]$
 for $\gamma$ and $\delta_B$,
and measure
\begin{equation*}	
r_B \in [\rLoB, \rHiB].
\end{equation*}
Our results are in agreement with the
current world averages~\cite{PDG2008}.

To facilitate the combination of these measurements with the results
of our Dalitz plot analysis of $B^\pm\to DK^\pm$, $D\to \KS h^+h^-$
$(h=K,\ \pi)$~\cite{babar_dk_DALITZ_PRL}, we exclude the $D\to \KS\phi$, $\phi\to\kk$
channel from this analysis -- thus removing events selected also
in~\cite{babar_dk_DALITZ_PRL} -- and then determine
\begin{eqnarray*}
  \Acpm &=& \AcpmValNoKsPhi\pm\AcpmErrStatNoKsPhi\stat\pm\AcpmErrSystNoKsPhi\syst\,,\\
  \Rcpm &=& \phantom{-}\RcpmValNoKsPhi\pm\RcpmErrStatNoKsPhi\stat\pm\RcpmErrSystNoKsPhi\syst\,.
\end{eqnarray*}
For comparison with the results of the $B^\pm\to DK^\pm$,
$D\to \KS h^+h^-$ analyses,
which are expressed in terms of the variables $x_{\pm} =
r_B\cos(\delta_B\pm\gamma)$ and $y_{\pm}=r_B\sin(\delta_B\pm\gamma)$, 
we express our results for the GLW observables in terms of
$x_+$ and $x_-$. We measure
\begin{eqnarray*}
  x_+   &=&            \xpValNoKsPhi\pm\xpErrStatNoKsPhi\stat\pm\xpErrSystNoKsPhi\syst\,,\\
  x_-   &=& \phantom{-}\xmValNoKsPhi\pm\xmErrStatNoKsPhi\stat\pm\xmErrSystNoKsPhi\syst\,,
\end{eqnarray*}
at 68\% CL. These results are in good agreement with
the current world averages~\cite{HFAG} and have precision comparable to
the single most precise measurements~\cite{babar_dk_DALITZ_PRL}.
We also evaluate $r_B$ after the exclusion of the $D\to \KS\phi$ channel,
and obtain a weak constraint on $r_B^2 = x_\pm^2\pm y_\pm^2$:
\begin{equation*}
r_B^2 = \rbSqValNoKsPhi\pm\rbSqErrStatNoKsPhi\stat\pm\rbSqErrSystNoKsPhi\syst
\end{equation*}
at 68\% CL.

\section{Acknowledgements}
We are grateful for the 
extraordinary contributions of our \pep2\ colleagues in
achieving the excellent luminosity and machine conditions
that have made this work possible.
The success of this project also relies critically on the 
expertise and dedication of the computing organizations that 
support \babar.
The collaborating institutions wish to thank 
SLAC for its support and the kind hospitality extended to them. 
This work is supported by the
US Department of Energy
and National Science Foundation, the
Natural Sciences and Engineering Research Council (Canada),
the Commissariat \`a l'Energie Atomique and
Institut National de Physique Nucl\'eaire et de Physique des Particules
(France), the
Bundesministerium f\"ur Bildung und Forschung and
Deutsche Forschungsgemeinschaft
(Germany), the
Istituto Nazionale di Fisica Nucleare (Italy),
the Foundation for Fundamental Research on Matter (The Netherlands),
the Research Council of Norway, the
Ministry of Education and Science of the Russian Federation, 
Ministerio de Ciencia e Innovaci\'on (Spain), and the
Science and Technology Facilities Council (United Kingdom).
Individuals have received support from 
the Marie-Curie IEF program (European Union), the A. P. Sloan Foundation (USA) 
and the Binational Science Foundation (USA-Israel).

\bibliography{Bibliography}{}
\bibliographystyle{h-physrev5}

\begin{table*}[!ht]
\center
\caption{Fit result of the three final fits to data, before correcting
for fit biases (see Section~\ref{sec:systematics}).}
\label{tab:fitresult}
\begin{tabular}{l|cc|ccc|c}
\hline\hline\\[-2.5ex]

Parameter & \kknc& \pipinc& \kspiznc& \ksomeganc& \ksphinc& \kpinc\\
\hline
             $A_{\CP}^{\sigk}$ & \multicolumn{2}{c|}{$    0.242 \pm 0.065    $}     & \multicolumn{3}{c|}{$   -0.089 \pm 0.066	 $}				    & $	  -0.008\pm0.022    $ \\
             $A_{\CP}^{\sigp}$ & \multicolumn{2}{c|}{$    0.003 \pm 0.015    $}     & \multicolumn{3}{c|}{$   -0.009 \pm 0.014	 $}   				    & $  -0.0116\pm0.0050   $ \\
                        $\Rkp$ & \multicolumn{2}{c|}{$   0.0897 \pm 0.0063   $}     & \multicolumn{3}{c|}{$   0.0808 \pm 0.0056  $}                                 & $   0.0753\pm0.0018   $ \\
                           $m$ & \multicolumn{2}{c|}{$   0.0204 \pm 0.0030   $}     & \multicolumn{3}{c|}{$   0.0206 \pm 0.0029  $}                                 & $  0.02143\pm0.00089  $ \\
\hline
             $A_{\CP,f}^{\BB}$ & $   -0.004\pm0.045    $  & $   -0.043\pm0.047    $ & n/a                      & n/a                      & n/a                     & $   -0.043\pm0.017    $ \\
             $A_{\CP,p}^{\qq}$ & $    0.012\pm0.016    $  & $   -0.016\pm0.018    $ & $   -0.002\pm0.011    $  & $    0.012\pm0.020    $  & $	-0.069\pm0.060    $ & $   -0.027\pm0.016    $ \\
             $A_{\CP,f}^{\qq}$ & $   -0.004\pm0.011    $  & $  -0.0044\pm0.0098   $ & $   0.0021\pm0.0071   $  & $   -0.004\pm0.013    $  & $	 0.001\pm0.039    $ & $  -0.0016\pm0.0068   $ \\
             $f^{\qq}_{\fshr}$ & $    0.326\pm0.026    $  & $ 0.49$ (fixed) 	    & $    0.520\pm0.030    $  & $ 0.27$ (fixed)          &        n/a 	            & $    0.396\pm0.018    $ \\
    $\sigma^{\qq}_{\fshr,l,1}$ & $    0.160\pm0.016    $  & $    0.258\pm0.023    $ & $    0.206\pm0.014    $  & $    0.175\pm0.034    $  & $	0.2758\pm0.0092   $ & $    0.198\pm0.014    $ \\
    $\sigma^{\qq}_{\fshr,l,2}$ & $   0.1742\pm0.0020   $  & $   0.2047\pm0.0024   $ & $   0.1546\pm0.0015   $  & $   0.1963\pm0.0028   $  &	   n/a   	    & $   0.1965\pm0.0017   $ \\
    $\sigma^{\qq}_{\fshr,r,1}$ & $    0.312\pm0.011    $  & $    0.329\pm0.011    $ & $   0.3541\pm0.0068   $  & $    0.317\pm0.019    $  & $	 0.447\pm0.014    $ & $   0.3068\pm0.0061   $ \\
    $\sigma^{\qq}_{\fshr,r,2}$ & $    0.231\pm0.014    $  & $    0.268\pm0.018    $ & $    0.275\pm0.020    $  & $    0.238\pm0.013    $  &	   n/a  	    & $    0.237\pm0.010    $ \\
     $\sigma_{\de,\mes}^{\BB}$ & n/a                      & n/a                     & n/a                      & n/a                      & n/a                     & $  0.01048\pm0.00057  $ \\
               $a_{\de}^{\qq}$ & $    -0.96\pm0.14     $  & $    -0.71\pm0.14     $ & $   -0.924\pm0.099    $  & $    -1.04\pm0.18     $  & $-0.48$ (fixed) 	    & $    -0.88\pm0.10     $ \\
                   $\mu_{\de}$ & $    -2.62\pm0.32     $  & $    -1.36\pm0.57     $ & $    -1.80\pm0.35     $  & $    -2.87\pm0.59     $  & $	 -0.95\pm0.75	  $ & $   -1.527\pm0.092    $ \\ 
                $\sigma_{\de}$ & $    16.63\pm0.27     $  & $    14.82\pm0.49     $ & $    17.01\pm0.29     $  & $    16.10\pm0.52     $  & $	 15.82\pm0.60	  $ & $   15.424\pm0.076    $ \\ 
                  $\mu_{\mes}$ & $  5278.56\pm0.12     $  & $  5278.61\pm0.20     $ & $  5278.62\pm0.12     $  & $  5278.50\pm0.22     $  & $  5278.99\pm0.25	  $ & $ 5278.586\pm0.033    $ \\ 
             $\sigma_{\mes,l}$ & $    2.207\pm0.081    $  & $     2.12\pm0.15     $ & $    2.299\pm0.084    $  & $     2.11\pm0.16     $  & $	  2.33\pm0.17	  $ & $    2.210\pm0.022    $ \\ 
             $\sigma_{\mes,r}$ & $    2.897\pm0.081    $  & $     2.83\pm0.15     $ & $    2.922\pm0.084    $  & $     3.08\pm0.16     $  & $	  2.72\pm0.17	  $ & $    2.852\pm0.023    $ \\ 
                 $N_{p}^{\BB}$ & $       79\pm29       $  & $      346\pm52       $ & $      176\pm43	    $  & $	180\pm48       $  & $ 3$ (fixed) 	    & $      328\pm40	    $ \\
                 $N_{f}^{\BB}$ & $     1430\pm82       $  & $     1517\pm142      $ & $     1930\pm102      $  & $     1195\pm109      $  & $	   119\pm20	  $ & $     7717\pm170      $ \\
                 $N_{p}^{\qq}$ & $     4005\pm69       $  & $     3456\pm76       $ & $     8587\pm101      $  & $     2675\pm68       $  & $	   284\pm17	  $ & $     4722\pm77	    $ \\
                 $N_{f}^{\qq}$ & $    10890\pm125      $  & $    13019\pm176      $ & $    21657\pm172      $  & $     6673\pm124      $  & $	   716\pm29	  $ & $    28007\pm205      $ \\
         $N^{\sigp}_{\rm tot}$ & $     4091\pm70       $  & $     1230\pm41       $ & $     4182\pm73	    $  & $     1440\pm45       $  & $	   648\pm27	  $ & $    44631\pm232      $ \\
\hline\hline
\end{tabular}
\end{table*}

\end{document}